\documentclass[manuscript]{acmart}
\AtBeginDocument{%
  \providecommand\BibTeX{{%
    \normalfont B\kern-0.5em{\scshape i\kern-0.25em b}\kern-0.8em\TeX}}}
    
\usepackage{enumerate}
\usepackage{booktabs}
\usepackage{tabularx}
\usepackage{tabularray}
\usepackage[normalem]{ulem}
\usepackage{soul}
\setcopyright{acmcopyright}
\copyrightyear{2024}
\acmYear{2024}
\acmDOI{XXXXXXX.XXXXXXX}

\acmConference[OzCHI]{OzCHI 2024 the 36th Australian Conference on Human-Computer Interaction}{December 2024}{Brisbane, Queensland, Australia}
\acmBooktitle{OzCHI 2024 the 36th Australian Conference on Human-Computer Interaction,
 11.30--12.04, 2024, Brisbane, Australia} 
\acmPrice{}
\acmISBN{}

\begin{document}

% \title{Scaling Up Co-Design with Chatbot Facilitators}
\title{Facilitating Asynchronous Idea Generation and Selection with Chatbots}

\author{Joongi Shin}
\affiliation{%
  \institution{Aalto University}
  \city{Espoo}
  \country{Finland}}
\email{joongi.shin@aalto.fi}

\author{Ankit Khatri}
\affiliation{%
  \institution{Aalto University}
  \city{Espoo}
  \country{Finland}}
\affiliation{%
  \institution{NIT Jalandhar}
  \city{Jalandhar}
  \country{India}}
\email{ankitk.cs.21@nitj.ac.in}

\author{Michael A. Hedderich}
\affiliation{%
  \institution{LMU Munich \& MCML}
  \city{Munich}
  \country{Germany}}
\email{Michael.Hedderich@lmu.de}

\author{Andrés Lucero}
\affiliation{%
  \institution{Aalto University}
  \city{Espoo}
  \country{Finland}}
\email{lucero@acm.org}

\author{Antti Oulasvirta}
\affiliation{%
  \institution{Aalto University}
  \city{Espoo}
  \country{Finland}}
\email{antti.oulasvirta@aalto.fi}
\renewcommand{\shortauthors}{}
\newcommand{\ci}[1]{\textcolor{black}{#1}}

\begin{abstract}
People can generate high-quality ideas by building on each other's ideas. By enabling individuals to contribute their ideas at their own comfortable time and method (i.e., asynchronous ideation), they can deeply engage in ideation and improve idea quality. However, running asynchronous ideation faces a practical constraint. Whereas trained human facilitators are needed to guide effective idea exchange, they cannot be continuously available to engage with individuals joining at varying hours. In this paper, we ask how chatbots can be designed to facilitate asynchronous ideation. For this, we adopted the guidelines found in the literature about human facilitators and designed two chatbots: one provides a structured ideation process, and another adapts the ideation process to individuals' ideation performance. We invited 48 participants to generate and select ideas by interacting with one of our chatbots and invited an expert facilitator to review our chatbots. We found that both chatbots can guide users to build on each other's ideas and converge them into a few satisfying ideas. However, we also found the chatbots' limitations in social interaction with collaborators, which only human facilitators can provide. Accordingly, we conclude that chatbots can be promising facilitators of asynchronous ideation, but hybrid facilitation with human facilitators would be needed to address the social aspects of collaborative ideation.

\end{abstract}

\begin{CCSXML}
<ccs2012>
   <concept>
       <concept_id>10003120.10003130</concept_id>
       <concept_desc>Human-centered computing~Collaborative and social computing</concept_desc>
       <concept_significance>500</concept_significance>
       </concept>
 </ccs2012>
\end{CCSXML}

\ccsdesc[500]{Human-centered computing~Collaborative and social computing}

\keywords{Asynchronous ideation, conversational agent, facilitator}

% A "teaser" image appears between the author and affiliation
% information and the body of the document, and typically spans the
% page.
%   \includegraphics[width=\textwidth]{figures/conversation_flow.png}
%   \caption{The conversation flow of chatbot-facilitated idea generation and selection. The chatbot engage with individual users and guide them to actively build on each other's ideas and opinions.}
% %   \caption{** going to change it other conceptual model ** A conceptual model of idea generation without guidance (top) and with our chatbot facilitators (bottom). The chatbots aim to lead generating as many ideas as possible By showing other group members’ ideas (inspirations) and suggesting how users can build on the others‘ ideas (ideation methods).}
%   \Description{}
%   \label{fig:conversation_flow}

% \begin{figure}[b]
% \noindent\fbox{%
% \parbox{\dimexpr\linewidth-2\fboxsep-2\fboxrule\relax}{%
% \begin{tabular}{l}
% The word count of this paper is \textcolor{blue}{8568}. % appendix 1811
% \end{tabular}
% }%
% }
% \end{figure}

\maketitle

% \newcommand{\revision}[1]{\textcolor{blue}{#1}}
% % \newcommand{\revision}[1]{\textcolor{black}{#1}}
% \newcommand{\original}[1]{\textcolor{red}{#1}}

\section{Introduction}
\begin{figure}[b!]
  \centering
  \includegraphics[width=\linewidth]{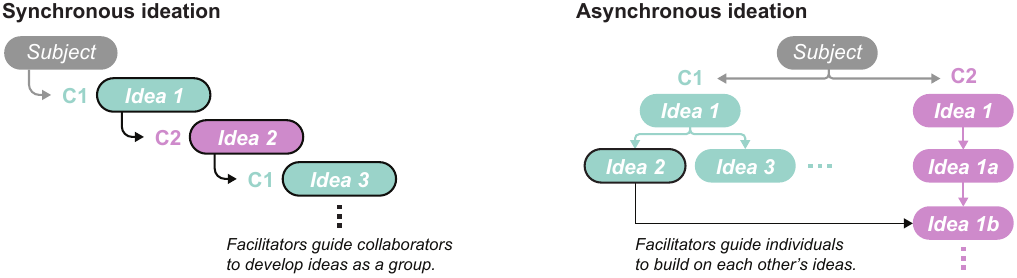}
  \caption{A conceptual model of synchronous and asynchronous ideation. Unlike synchronous ideation where collaborators (C1 and C2) develop ideas by taking turns, we focus on asynchronous ideation where collaborators generate ideas individually, at their own pace and method (right). For example, while C1 diversifies ideas, C2 could focus on improving ideas. Here, facilitators' role will be presenting C1's idea as an inspiration to C2.}
  \label{fig:ideation_type}
\end{figure}

\ci{Human creativity thrives on collective intelligence~\cite{surowiecki:book:2004}. By exchanging ideas and perspectives, people can diversify and converge to innovative solutions beyond individuals' consideration~\cite{kohn:jesp:2011,casakin:jdci:2021,brown:cdps:2002,surowiecki:book:2004}. One form of collaborative ideation is \emph{asynchronous ideation}, wherein collaborators independently contribute ideas through a shared online platform
% --- at their own comfortable time and pace
~\cite{lauren:2020:asynchronous,yossi:2021:asyncBrainstorming,siangliulue:cscw:2015}. In contrast to synchronous ideation, where collaborators ideate together in real time, asynchronous ideation offers flexibility for individuals to ideate at their own comfortable pace and method
% removing temporal barriers between collaborators 
(Figure~\ref{fig:ideation_type}). This enables individuals to reflect deeply on their ideation process, which has been shown to improve idea quality~\cite{davis:sdrj:2021}.
% Enabling effective asynchronous ideation could benefit diverse creative and societal endeavors~\cite{chan:2016:largeScale,fujita:2017:largeScale,siangliulue:2016:largeScale,sanders:codesign:2008}, such as involving a large number of stakeholders in the design process (i.e., co-design)~\cite{warren:2022:asyncCodesign} and generating petition ideas with a population of citizens~\cite{andreas:2010:petition}.
}

\ci{However, facilitating effective idea exchange in asynchronous settings faces a practical challenge; that is, human \emph{facilitators} cannot be available continuously to guide individuals joining at varying hours~\cite{michinov:2005:asyncBrainstorming}. In any collaborative ideation events, trained facilitators take crucial roles in guiding collaborators to \emph{build on each other's ideas}~\cite{gaudio:codesign:2020,lee:codesign:2008}. They actively engage with individuals and lead them to generate, criticize, refine, and select a few promising ideas as a group effort.}
% Throughout the process, facilitators focus collaborators on ideation goals, provide inspiration, and help them connect their ideas.}
Without facilitators, collaborators may ideate ineffectively, diverging towards unrelated subjects and failing to reach a consensus~\cite{vaajakallio:idc:2009,lee:codesign:2008}.

\begin{figure}[b!]
  \centering
  \includegraphics[width=\linewidth]{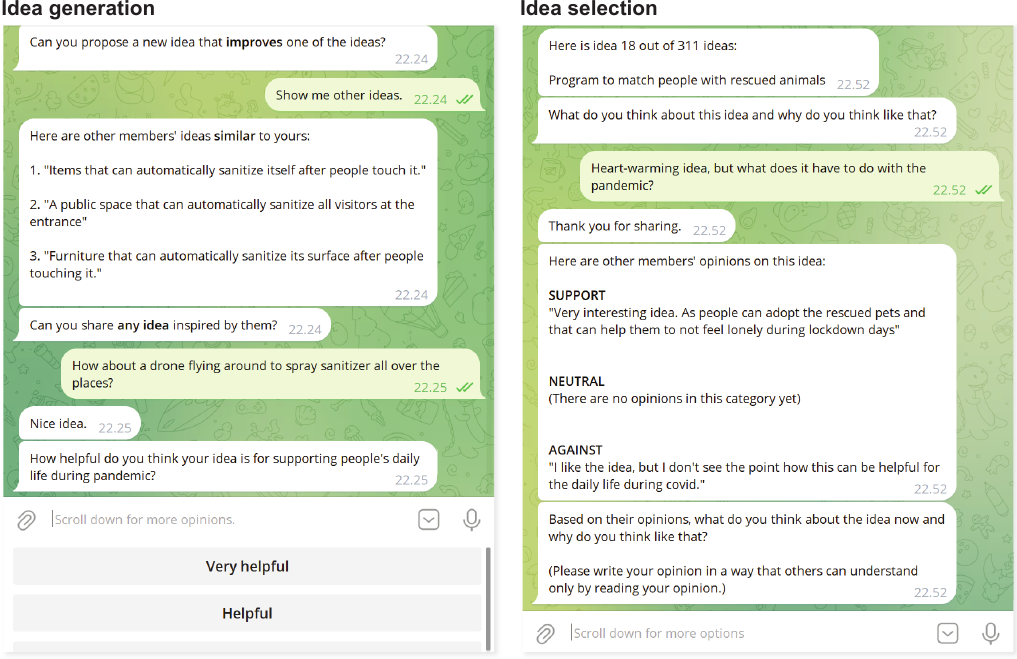}
  \caption{\ci{We designed chatbots that interact with individual collaborators and facilitate their asynchronous idea generation (left) and selection (right). We adapted guidelines from the literature on human facilitators to design our chatbots' behaviors, presenting other collaborators' ideas as inspiration and suggesting ideation methods to guide individuals to build on each other's ideas.}}
  \label{fig:apparatus}
\end{figure}

In this paper, we ask \emph{how to design conversational agents to facilitate asynchronous idea generation and selection} (Figure~\ref{fig:apparatus}).
Unlike human facilitators, conversational agents such as chatbots have no restriction on being continuously available, which could provide facilitation on demand.
While prior arts have shown chatbots' potential in facilitating online discussion and brainstorming \cite{hadfi:2021:onlineDiscussion, Ito:2022:onlineDiscussion, kim:cscw:2021, shin:chiea:2021, lavric:2023:chatbotFacilitator}, they focus on the synchronous discussion as a group (e.g., collaborators developing ideas by taking turns) and do not investigate asynchronous ideation (e.g., collaborators building on each other's ideas independently).
% Prior arts have shown chatbots' potential in facilitating ongoing discussion and brainstorming~\cite{kim:cscw:2021,shin:chiea:2021,lavric:2023:chatbotFacilitator}. 
% Yet, no studies have investigated how chatbots could guide individuals to build on each other's ideas in asynchronous settings.
Accordingly, we hypothesize that chatbots can adopt human facilitators' behaviors for \emph{converging individuals' ideation efforts into a collaborative effort} (Figure~\ref{fig:conceptual_model}). 
Chatbots could;

\begin{itemize}
    \item present other collaborators' ideas and opinions as \emph{inspirations};
    \item suggest \emph{ideation methods} for building on the others' ideas and opinions; and
    \item request individuals to \emph{rate} their ideas considering their \ci{ideation} goal.
\end{itemize}

\ci{Building chatbots has become more accessible thanks to Large Language Models (LLMs), but it is unclear how to design the interactions so that they can actually facilitate asynchronous ideation. To close this gap, we distill the guidelines found in the literature on human facilitators~\cite{puccio:jcb:2020,paulus:fp:2018,thoring:desire:2011} and propose two chatbots: One provides a structured ideation process (i.e., \textbf{structured facilitator}) and another adapt the ideation process to individuals' ideation performance (i.e., \textbf{adaptive facilitator}).}

\ci{The structured facilitator guides collaborators to first diversify ideas and then improve them. This approach has been proven to be effective in generating a large number of ideas, which also increases the number of high-quality ideas~\cite{paulus:jcb:2011}. During the idea selection phase, \ci{the facilitator} focuses users' attention on the ideas that have been rated to be helpful~\cite{dahl:chi:2022}.}
We designed the \ci{adaptive} facilitator inspired by how human facilitators spontaneously \ci{alter their guidance to provide tailored ideation processes to individuals~\cite{dahl:chi:2022}. The adaptive facilitator} provides inspirations (i.e., similar or dissimilar ideas) and ideation methods (i.e., generate any or improved ideas) based on how well individuals generate ideas from them. During the idea selection phase, the facilitator focuses users' attention on the ideas that have `uncertain' group opinions (i.e., ideas with fewer and more diverse opinions). 
\ci{By studying both approaches, we demonstrate how human facilitators' behaviors can be translated into chatbots.}

The chatbots are supported by a semantic similarity classifier that can \ci{retrieve and present} other collaborators' ideas by their similarity to individuals' own ideas.
For this, we tested a prompt-based classifier using LLM (GPT-4)~\cite{gatto:2023:stsGPT} and a fine-tuned classifier using a pretrained language model (DistilBERT)~\cite{Chandrasekaran:2021:stsdistilBERT}). We found that the fine-tuned classifier can estimate the semantic similarity as accurately as the prompt-based classifier but at a faster processing speed, making it more suitable for conversational interaction with collaborators.
In addition, we make the adaptive facilitator provide helpful inspirations and ideation methods based on how well users can generate quality ideas from them. For this, we used the Multi-Armed Bandit (MAB) algorithm~\cite{rl_book}.
% MAB is a probabilistic inference approach for finding a system's most rewarding behavior by exploring and exploiting its alternative behaviors~\cite{rl_book}. Our adaptive facilitator finds the inspirations and ideation methods for which individuals can generate more satisfying ideas with regard to the ideation goal. It also finds uncertain ideas that require more evaluation from other group members.

We conducted two studies to understand \ci{the strengths and weaknesses of our chatbots in facilitating asynchronous ideation. First, we assessed the chatbots from collaborators' perspectives. We invited 48 participants to generate and select ideas by interacting with one of our chatbots. We analyzed their interaction behaviors, perception of ideating through the chatbots, and satisfaction with the resulting ideas.
Then, we explored our chatbots' potential with a human facilitator, an expert who provided critical perspectives based on their facilitation practice over 25 years. We sensitized the expert by having them facilitate an in-person ideation workshop in the same structure as our chatbots facilitated and interviewed their perspectives on facilitating asynchronous ideation with chatbots.}

The results show that both chatbots were helpful in building on other group members' ideas.
\ci{The structured facilitator was more helpful in diversifying ideas, and the adaptive facilitator could elicit satisfaction at a similar level to the human-facilitated ideation.}
% we observed that all participants had multiple ideation sessions at their own time and increased the pool of ideas as collective effort. Their satisfaction with selected ideas was also similar between the groups.
However, we also identified the potential limitations of chatbots \ci{such as the lack of accountability and social interaction that only human facilitators can provide. Accordingly, we conclude that chatbots can facilitate collaborative ideation in asynchronous settings, yet human experts' intermediate facilitation would still be required.}
\ci{With this paper, we uncover the positives and negatives of chatbots in facilitating asynchronous ideation (Table~\ref{tab:summary}) and make the following contributions}:
\begin{enumerate}
    \item We demonstrate designing two chatbot facilitators. Both chatbots adopt the literature-based guidelines for facilitating effective collaborative ideation and employ computational methods.
    \item We found the positives and negatives of the chatbot facilitators from collaborators' and an expert human facilitator's perspectives.
    \item We provide the implications of chatbot facilitators in practice. We also offer our code for implementing chatbot facilitators using our computational methods as well as sample prompts to support those who aim to implement LLM-based facilitators using our design guidelines~\footnote{Removed for anonymity}.
\end{enumerate}

% \begin{figure*}[h]
%   \centering
%   \includegraphics[width=\linewidth]{figures/paperScope.png}
%   \caption{\revision{The stages of a co-design process~\cite{lucero2012, thoring:desire:2011}. Our scope is facilitating collaborative ideation with chatbots in the early stages of co-design.}}
%   \Description{}
%   \label{fig:paperScope}
% \end{figure*}

% In co-design, the final outcome affects you. That is not chatbots to consider. Where (stakeholders) come from. Especially in co-design, facilitators help stakeholders understand each other's perspectives.
% synchronous: Pro: team building, gather ideas quickly, intense and focused sessions. Con: scheduling, group dynamics
% Asynchronous: Pro: no scheduling, ideation at own comfortable pace, Con: practical barriers for effective facilitation. 

\section{Related work}
Facilitators' roles in collaborative ideation have been extensively investigated in design and HCI literature~\cite{galabo:digest:2020,lee:codesign:2008,lucero2012,dahl:chi:2022,gaudio:codesign:2020}, which help train human facilitators. We build on such knowledge and propose an alternative way of facilitating \ci{asynchronous ideation with conversational agents.}

% Co-design projects often involve a large number of stakeholders. \revision{In such cases, prior approaches have been dividing stakeholders into smaller groups} or into multiple workshops that human facilitators can manage (Table~\ref{tab:num_codesign}). Accordingly, the process can be costly, needing more facilitators for a longer duration.
%
% \revision{To involve a large number of participants, scholars have investigated digitizing co-design activities~\cite{dalsgaard:nordichi:2022} and performing co-design workshops online~\cite{walsh:idc:2012}}. Still, these methods require facilitators.
%
% \dis{With this paper, we study artificial facilitators that could increase the scale of co-design.}

\subsection{Facilitators in collaborative ideation}
% What facilitators can do?
Regardless of individuals' expertise in collaborative activities, facilitators take essential roles by closely engaging with participants~\cite{galabo:digest:2020,lee:codesign:2008,lucero2012}.
% Facilitators' roles
% such as a trust builder, enabler, inquirer, direction setter, value provider, and users' advocate. Each role is responsible for enabling 
Dahl and Sharma \cite{dahl:chi:2022} identified six roles of facilitators that can enable effective idea exchange among group members, despite their varying communication skills or authority in design projects.
% Brief descriptions
% For instance, the trust builder helps participants to build trust among themselves and motivates all members to openly discuss their ideas and perspectives. The inquirer identifies the conflicts among group members' needs while challenging and interrogating group members to realize their true needs.
% 
In particular, we focus on the roles of enabler, direction setter, and users' advocate \ci{that provide direct support on ideation.} The enabler assists individual group members to actively participate in their discussion, lowering the barrier of sharing their ideas and perspectives. The direction setter aligns group members' collaborative efforts with their project goals. The users' advocate represents and negotiates each user group's perspective, convincing other group members to take each other's viewpoints.
These roles would be particularly important \ci{in asynchronous settings, where collaborators need to consider other group members' ideas and opinions despite having no direct conversation with each other.}
One remark is, all the above-mentioned facilitation is achieved through conversation. 

%"Social influence may cause individuals to modify their opinions, attitude and behaviors to be similar to the other they are interacting with~\cite{myers:press:1982} cited by~\cite{singh:dcc:2020}"

% For instance, facilitators encourage users to interact with each other, identify less activity users to contribution, 

% balance the communication among users with different social hierarchy, provide explicit actions that users can take when they build on each other's contributions~\cite{}, 
% 
% For instance, they mediate the communication among user groups in dispute, moderate the pace of discussion, or grouping users who can best ideate with each other, hence making users to focus on exchanging their ideas~\cite{}.

% Bad things happening without facilitators 
% While exchanging ideas, users can easily lose their focus on the design goal, diverge into unrelated design subjects, or do not realize how to converge their thoughts as a group~\cite{galabo:digest:2020, lee:codesign:2008, lucero2012}.

\subsection{Interactive support for collaborative ideation}
To generate and select ideas as a group, collaborators need a setting where they can effectively exchange their ideas.
% offline interactive systems 
In response, diverse interactive systems have been explored to support communication among a user group.
For instance, Mazalke et al. \cite{mazalke:tei:2009} developed a digital tabletop where users can spatially share their stories in person. 
% Walsh et al.~\cite{walsh:idc:2016} investigated co-design with children in a digital world where children could easily share their ideas. 
% 
Kennedy et al. \cite{kennedy:ijerph:2021} explored conducting collaborative design in a digital environment and identified that online meeting platforms enabled more balanced idea exchange among the collaborators.
Despite the utility of their systems, these approaches require users to collaborate at the same time, synchronously.
% Our novelty.
In contrast, we aim to enable users to collaborate asynchronously, by adapting the guidelines on facilitating effective idea generation~\cite{knoll:hicss:2010,faste:chi:2013}.

Similar to our interests, digital platforms have been investigated to enable asynchronous idea generation~\cite{klemmer:uist:2001,lucero2012,moghaddam:interact:2011}.
For instance, Siangliulue et al., \cite{siangliulue:cscw:2015} studied how users group multiple ideas by their semantic similarity and developed a platform where they contribute their ideas by the groups.
Online platforms such as Miro\footnote{\url{https://miro.com/}} and Mural\footnote{\url{https://www.mural.co/}} provide diverse functionalities for representing and organizing ideas.
% Unehara et al. developed a design-support system that generates alternative designs based on group members' vote on the initial set of designs~\cite{unehara:isais:2017}.
%
% Idea selection
% Several of them focus more on converging users' opinions as a follow-up of idea generation. Grünbacher and Briggs~\cite{grunbacher:ahicsc:2001} studied a shared platform where users can vote for the winning ideas and guide later negotiation. Moghaddam et al.~\cite{moghaddam:interact:2011} developed a discussion platform where users can track the alternative ideas and arguments among group members. 
% 
Online petition websites can also support a population to raise socially immanent issues in discussion~\cite{petition:}.
% 
% The aforementioned systems could enable a large user group to jointly discuss and converge their ideas.
% 
Most systems, however, expect collaborators to be self-motivated to discuss their ideas and \ci{do not provide adaptive guidance as human facilitators do}. Without facilitation, users would miss the others' meaningful contributions to build on or diverge ideas to unrelated subjects~\cite{vaajakallio:idc:2009,lee:codesign:2008}.
% Our novelty.
In response, we study artificial facilitators that can provide concrete guidance to individual collaborators in asynchronous settings.

\subsection{Chatbots as facilitators}
The recent advancements in LLMs have created powerful chatbots (e.g., ChatGPT~\footnote{\url{https://openai.com/blog/chatgpt}}) that can conduct human-like conversation~\cite{jakesh:pnas:2023}. 
Such LLM-based chatbots have been shown to support people's ideation, such as summarizing extensive text~\cite{nlp/he-etal-2023-z} and generating ideas~\cite{chatgpt/electronics12163535}. 
However, most applications have been limited to the interaction between a single user and a chatbot~\cite{chatgpt/kocaballi2023conversational,chatgpt/lamichhane2023evaluation,chatgpt/di:chi:2022,chatgpt/yan:iui:2022}.
% Despite LLM-based chatbots' competence, `being non-human' introduces challenges when integrating AI in collaboration between humans~\cite{aiFacilitate:reddy:2019, shin:chiea:2023, aiCollab:wang:chiea:2020}. In other words, chatbots should be designed to facilitate. Accordingly, we study how to design chatbots as the facilitators of collaborative ideation.
\ci{How to design chatbots (or instruct LLM-based chatbots) that facilitate collaborative ideation among people should be studied. In this paper, we demonstrate designing chatbot facilitators by adopting human facilitators' behaviors.}

% Little background
Diverse collaborative ideation methods (e.g., discussing ideas using physical probes~\cite{pernille:codesign:2021,kim:dis:2006}) have been practiced for users to explore and elaborate their ideas~\cite{kankainen:bit:2012,xie:bit:2012,sleeswijk:codesign:2005,kensing:cacm:1993,giaccardi:dis:2012}. All methods, however, are designed to be led by trained facilitators. Conducting collaborative ideation hence highly depends on facilitators' expertise and availability. 
In response, scholars have investigated diverse roles of conversational agents to support collaborative activities~\cite{park:cscw:2021,haqbeen:ci:2020,walsh:cscw:2019,hadfi:jssse:2021,seering:chi:2019,goda:ijis:2014}.
% 
% For instance, Goda et al.~\cite{goda:ijis:2014} investigated a chatbot as ideation partner that users can individually refine their own thoughts before discussing in groups. 
For instance, Wambsganss et al. \cite{wambsganss:chi:2021} developed a chatbot that guides users to improve their persuasive writing. These studies show a promising result that private discussion with a chatbot can prime users for collaboration.
In alignment with our aim, a few studies developed chatbots to facilitate ideation among group members. Kim et al. \cite{kim:chi:2020} developed a chatbot that moderates discussion in real-time, pacing collaborators' discussion, identifying inactive collaborators, and organizing their discussion points. The authors also investigated how chatbots can facilitate idea convergence at the end of discussion~\cite{kim:cscw:2021}.
Hadfi et al. \cite{hadfi:2021:onlineDiscussion} designed chatbots to facilitate discussion on an online forum, which guides participants to sequentially comment on the previous opinion. While these studies show the potential of facilitating collaborative ideation via chatbots when human facilitators are not available, they are limited to synchronous ideation where collaborators need to exchange ideas in turn.
% Lee et al.~\cite{lee:chi:2020} designed a platform that can assist facilitators in moderating an asynchronous discussion among collaborators.
% 
The most similar study to ours was conducted by Shin et al. \cite{shin:uist:2022}, developing a rule-based chatbot that facilitates consensus-building to resolve conflicts by enabling independent exploration of different opinions.
% 
% These studies show the potential of facilitating collaborative ideation via chatbots when human facilitators are not available.
% 
\ci{We extend the study by designing chatbot facilitators that can guide asynchronous idea generation and selection.}
\section{Literature-based guidelines for chatbot facilitation}
\begin{figure}[t!]
  \centering
  \includegraphics[width=\linewidth]{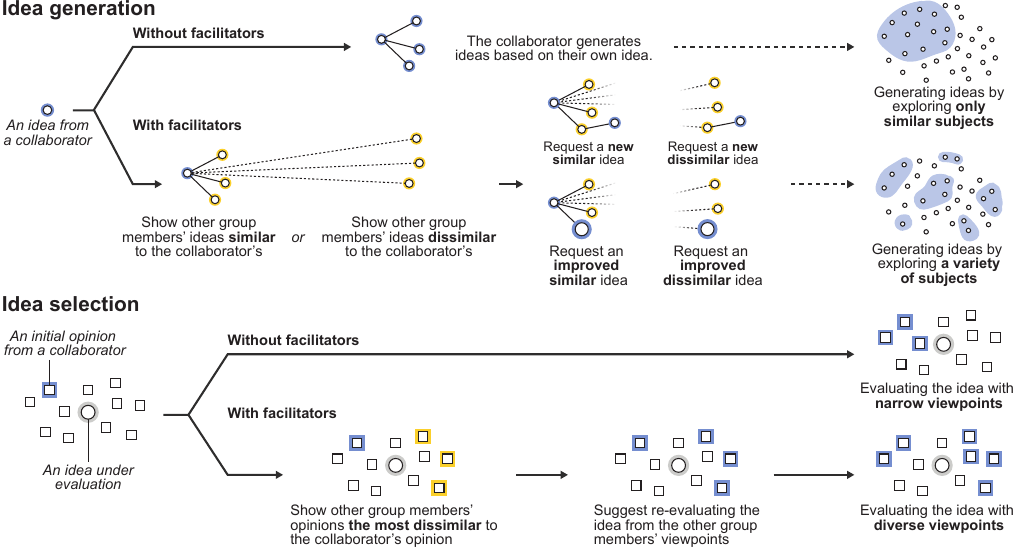}
  \caption{Collaborative idea generation (top) and selection (down) led by facilitators. By showing other group members' ideas and opinions, facilitators can guide individual collaborators to effectively generate more ideas and evaluate the ideas with more diverse viewpoints. We designed chatbot facilitators that could guide such effective ideation in asynchronous settings.}
  \label{fig:conceptual_model}
\end{figure}

We distilled guidelines found in the literature that demonstrate human facilitators' roles from empirical analysis, case studies, and interviews on facilitating collaborative design projects~\cite{lucero2012, gaudio:codesign:2020,lee:codesign:2008,dahl:chi:2022}. The guidelines include approaches for structuring group ideation and adaptively engaging with collaborators, boosting their collective effort in generating and selecting ideas. Accordingly, we made a conceptual model of effective facilitation (Figure~\ref{fig:conceptual_model}) and made four design decisions commonly applied to our chatbot facilitators (Table~\ref{tab:design_guideline}).

\begin{table}[b]
    \centering
    \small
    \begin{tabularx}{\textwidth}{X X}
    \toprule
    Literature-based guidelines & Our design decisions for chatbot facilitators \\ 
    \midrule
    Enable independent ideation to boost idea generation~\cite{paulus:fp:2018} & Engage with individual collaborators in isolated settings \\
    Present similar / dissimilar ideas to leverage group ideation~\cite{siangliulue:cscw:2015} & Provide inspirational ideas by their similarity \\
    Provide instructions for building on each other's ideas~\cite{lucero2012} & Suggest ideation methods \\
    Focus collaborators' attention on their ideation goal~\cite{dahl:chi:2022} & Request collaborators to rate their ideas against the ideation goal\\ 
    \bottomrule
    \end{tabularx}
    \caption{Exemplary literature-based guidelines we adopted to design our chatbot facilitators.}
    \label{tab:design_guideline}
\end{table}

\textit{First, chatbots should engage with individuals in isolated settings.}
\ci{Facilitators often divide collaborators to} work alone, without interfering with each other (e.g., Nominal group technique~\cite{nominalGroup}). This enables individuals to generate ideas without taking turns (i.e., production blocking) and without worrying about the others' criticism (i.e., evaluation apprehension), hence effectively increasing the number of ideas~\cite{knoll:hicss:2010,faste:chi:2013}.
\ci{In the process, facilitators pay close attention to individuals~\cite{dahl:chi:2022}, fostering their independent creativity~\cite{choi:ijsm:2020,paulus:fp:2018}.} 
We expect chatbot facilitators also to guide individuals in solitary settings (e.g., chatrooms dedicated to individuals), enabling ideation at their own pace and creativity.

\textit{Second, chatbots should present inspirational ideas by their similarity to the proposed ideas.}
Facilitating collaborators to combine common ideas can lead to generating impactful ideas~\cite{kohn:jesp:2011}.
Showing other group members' dissimilar ideas can lead to generating unique ideas~\cite{siangliulue:cscw:2015}.
Likewise, we expect showing similar ideas to inspire exploration in the related subjects or different subjects to avoid duplicates.
% and generating more improved ideas. 
By showing dissimilar ideas, we expect collaborators to explore the subjects that they would not have considered otherwise~\cite{paulus:fp:2018,brown:cdps:2002}.
% IS?
We apply the same reasoning for idea selection: showing dissimilar opinions would make collaborators consider diverse viewpoints and make correct judgments.

% Show only three ideas:
% -> Minimize overloading users with potential information and focus their attention.
% -> Reduce the change of showing false-postives (i.e., intend to show similar ideas but the ideas are not similar ideas). NLP model is not perfect. A few study shows that users also selectively respond to the most correct ones when artificial agents present multiple options.

\textit{Third, chatbots should suggest ideation methods.}
Facilitators present a set of actions that collaborators can build on each others' contributions~\cite{lucero2012,gaudio:codesign:2020}. For instance, during group brainstorming, facilitators specify the ideation method that group members can use to improve each other's ideas (e.g., ``Imagine why this idea will not work'' and ``Imagine what would be needed to make this idea work'')~\cite{roundRobin}.
% IG
We adopt the behaviors and design the chatbots to suggest ideation methods in addition to providing inspiration.
% users to propose new or improved ideas.
They suggest collaborators propose new or improved ideas during idea generation and re-evaluate the shown ideas from the others' viewpoint (i.e., perspective-taking) during idea selection.
% 
% In particular, does not bring more insights to the users' initial opinion, thinking about the others' viewpoints may help them build consensus in the later phases of co-design~\cite{coleman:book:2014, susskind:book:1999}.

\textit{Lastly, chatbots should request collaborators to rate their ideas regarding their ideation goals.}
\ci{While interacting with proposed ideas and opinions, collaborators could lose their focus and start developing unrelated ideas.
To resolve this, facilitators often interfere and redirect collaborators' attention to the ideation goal~\cite{dahl:chi:2022}.} Similarly, we employ the rating approach, which could quickly form group opinions and focus collaborators' attention on the ideation goal.
% Voting has been used in many consensus-building processes~\cite{coleman:book:2014, susskind:book:1999}.
% Especially when a large number of users need to reach consensus
% , where group members vote for the most interesting ideas or rate the usefulness of ideas
% This aligns well with our aim where large user groups need to form group opinions. 
% It is often followed by additional discussion, where group members' discuss their satisfaction to the selected ideas.
% 
We expect chatbots asking collaborators to rate their ideas with regard to ideation goals can help regain their focus after seeing or proposing unrelated ideas.

% \subsubsection{Promote generating any ideas}
% We build on the principle that generating many ideas can lead to generating creative ideas (i.e., quantity over quality)~\cite{}. The basic mechanism of this process is that... 
% \textit{Lastly, chatbots should define the order of ideas for collaborators to review.} As a result of idea generation, a large number of ideas could be generated: evaluating all ideas will be laborious. 
% % Whereas chatbots could filter out duplicated ideas based on NLP technologies, there could be similar and dissimilar ideas that are worth users' attention (henceforth, remarkable ideas). 
% Therefore, we design the chatbots to set ideas in the order that a group of individuals can reach a consensus without reviewing every idea. Each structured and adaptive facilitator solves this in a unique way, which we describe in the following subsections.

\section{Design of chatbot facilitators}
\begin{figure}[t!]
  \centering
  \includegraphics[width=\linewidth]{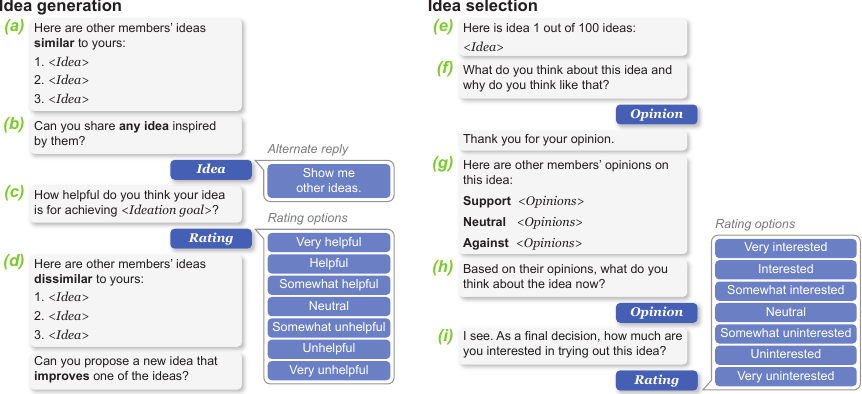}
  \caption{During idea generation (left), the chatbots repeat the cycle of presenting other group members' ideas as inspiration (a), suggesting an ideation method (b), and requesting ratings on the users' own ideas (c). Then, the chatbots continue showing other types of inspiration and ideation methods (d). During idea selection (right), the chatbots show a collected idea (e), request users' initial opinion (f), show other group members' opinions (g), suggest re-evaluating the idea (h), and request ratings on the idea.}
  \label{fig:conversation_flow}
\end{figure}

We designed the structured and adaptive facilitators that guide individual collaborators to generate and select ideas by building on each other's contributions. Overall, both chatbots share the same conversation flow as shown in Figure~\ref{fig:conversation_flow}. 

% Walkthrough IG
The chatbots begin the idea generation phase by introducing an ideation goal and how they will support the users' ideation. 
% 
% The chatbots then ask the users about their personal experiences related to the design subject (sensitizing activity)~\cite{kennedy:ijerph:2021}.
% 
% After the users reply, the chatbots introduce the design goal and how they will support the users' ideation.
% 
They also encourage users to propose as many ideas as possible, unrestricted by current technology or resources.
% Cycle
Afterward, the chatbots start the main cycle of idea generation.
% Show inspiration
(a) The chatbots first show either three similar or dissimilar ideas to the users' latest idea (they show the common or rare ideas if the users just began the interaction, hence have not yet shared any idea).
% Suggest ideation method
Then, (b) the chatbots suggest the users propose any or improved idea.
% Collect users' ideas.
% In response, the users share either a new or an improved idea.
% MAB related
% If the users cannot share any ideas based on the inspiration and ideation method, they can request the chatbot to show other ideas, which gives the lowest reward to the chatbot.
% 
% request rating
After the users share an idea, (c) the chatbots ask them to rate their own idea using a 7-point Likert scale, considering how well it achieves the ideation goal.
% Post rating
After the rating, (d) the chatbots repeat the three actions until the end of idea generation.

% IS
The chatbots begin the idea selection phase by reminding users about the ideation goal, highlighting the number of generated ideas, and encouraging users to review as many ideas as possible.
% 
% They also describe how they will assist users and inform that three ideas with the highest rating will be selected as the end result.
% 
% The chatbot encourages users to review as many ideas as possible, without forcing them to evaluate all the collected ideas.
% 
After the introduction, the chatbots begin the cycle of idea selection. (e) The chatbots present one idea at a time and (f) ask for the users' initial opinions on the idea. 
Then, (g) the chatbots present other group members' opinions on the idea that are most dissimilar to the users' initial opinion. 
The chatbots present them in three categories (support, neutral, and against) based on the others' final ratings of the idea.
(h) The chatbots then suggest users re-evaluate the idea considering the other group members' opinions. As a response, the users share either a new opinion or keep their initial one.
The chatbots thank users for sharing their opinions and (i) request the users to rate how much they would like to try out the idea using a 7-point Likert scale. After collecting the users' ratings, the chatbots repeat the three actions until the idea selection phase ends.

The structured and adaptive facilitators differ in how they decide which inspirations and ideation methods to present during idea generation and which ideas to focus users' attention on during idea selection. In particular, in the idea selection phase, we expect evaluating all generated ideas to be laborious. Therefore, we designed the chatbot facilitators to present ideas in the order that a group of individuals can reach a consensus without reviewing every idea. The following subsections provide details on the differences between the chatbot facilitators.
% % Whereas chatbots could filter out duplicated ideas based on NLP technologies, there could be similar and dissimilar ideas that are worth users' attention (henceforth, remarkable ideas). 
% Therefore, we design the chatbots to set ideas in the order that a group of individuals can reach a consensus without reviewing every idea. Each structured and adaptive facilitator solves this in a unique way, which we describe in the following subsections.

\subsection{Structured facilitator}
% (Figure~\ref{fig:rule_rl}).
% \begin{figure*}[h]
%   \centering
%   \includegraphics[width=\linewidth]{figures/rule_rl(new).png}
%   \caption{\ci{update the figure} Idea generation (left) and selection (right) led by the structured facilitator (top) and by the adaptive facilitator (bottom). The structured facilitator predefines the structure of ideation whereas the adaptive facilitator provides adaptive guidance responding to the individuals' ideation performance and progress as a group.}
%   \Description{}
%   \label{fig:rule_rl}
% \end{figure*}
\ci{The structured facilitator provides the same ideation process to all users. The facilitator divides idea generation into two parts: First diversifying ideas and then improving them.}
This is based on the principle that a group of users performs better when each ideation session has a single purpose, concentrating on one type of ideation at a time~\cite{puccio:jcb:2020,paulus:fp:2018,thoring:desire:2011}. \ci{This also aligns with the `quantity over quality' approach~\cite{paulus:jcb:2011}, which promotes generating a large number of ideas without worrying about their quality, hence exploring a broad spectrum of ideas.}
% 
% The basic concept is that generating as many ideas as possible without considering their quality can uncover a wide spectrum of design subjects, which leads to generating more creative ideas.
% 
\ci{Our facilitator shows dissimilar ideas and suggests proposing any ideas to assist the divergent thinking in the first part. To assist in improving ideas, it shows similar ideas and suggests improving one of them.}

% How does it structure IS?
\ci{During idea selection, the structured facilitator} presents ideas in the descending order of their ratings received during the idea generation phase. 
% 
% Ideas with the higher ratings get higher priority and will be reviewed before others.
% 
\ci{This strategy can focus collaborators' attention} on the ideas with higher potential, filtering out less interesting ones and reducing individuals' effort in reviewing all proposed ideas~\cite{litcanu:sbs:2015,knight:dj:2019,faste:chi:2013}.

\subsection{Adaptive facilitator}
% What it does for Idea generation
During idea generation, the adaptive facilitator selects inspirations and ideation methods based on users' performance in generating ideas.
For instance, if users generate more helpful ideas by responding to similar ideas and thinking about new ones, the adaptive facilitator will continue guiding the users with the same type of inspiration and ideation method.
If users start to generate less helpful ideas, then the facilitator will suggest different types of inspiration and ideation methods.
Whereas this approach will bring multiple types of ideation in a single phase and contradicts the principles applied to the structured facilitator, we hypothesize that the same principles may not apply when users can generate ideas at their own pace in the asynchronous setting. 
% rationale for adaptively showing actions
% For instance, if a group of users prefer improving ideas than generating ideas, forcing them to diversify ideas would be ineffective.
% 
% Since they collaborate asynchronously, 
% This can be more effective for them to generate ideas in a best way that they can.
% 
% For this, the bandit-based facilitator tries four actions (2 inpisrations X 2 ideation methods) and identifies with which action the users can produce more helpful ideas.

% What does it do for idea selection? 
During idea selection, the adaptive facilitator continuously updates which idea it should present next based on the `uncertainty' of group opinions on the ideas. 
We assume that, after collecting sufficient number of opinions and ratings on an idea, users can be certain whether they would like or dislike the idea as a group. In other words, collecting more opinions on such ideas will not add more information about how collaborators think about the idea.
Therefore, it would be more efficient if users review the other ideas that have not yet collected enough opinions.
This can lead to distributing users' effort, which would be especially useful when there are many ideas to review.
\section{Computational approaches for chatbot facilitators}
\ci{We implemented our chatbots' behaviors using Natural Language Processing (NLP) and probabilistic inference. This section describes the overview of applied methods. We provide detailed technical descriptions of the methods and their evaluations in Appendix \ref{appendix:ssc} and \ref{appendix:mab}.}

\subsection{Semantic similarity classifier}
Both chatbot facilitators present other group members' ideas and opinions by their similarity to the users' latest input.
\ci{For this, they need to perform a Semantic Textual Similarity (STS) prediction, measuring how closely related two sentences are when looking at the information they convey. In the context of facilitating ideation, we aim for chatbots that compute STS aligned with humans' perception of idea similarity and at an interaction rate of speed.}

% LLMs could perform STS better than smaller language models, whereas a classifier could perform STS faster than LLMs.
\ci{In a preliminary experiment, we evaluated two classifiers for this STS task: a \textbf{prompt-based classifier} using LLM and a \textbf{fine-tuned classifier} based on a pretrained language model.} For the prompt-based classifier, we used GPT-4\footnote{gpt-4-0613: https://platform.openai.com/docs/models/overview}, the latest model at the time of testing, and based our prompts on Gatto et al.'s instruction~\cite{gatto:2023:stsGPT}. For the fine-tuned classifier, we used DistilBERT~\cite{nlp/Sanh19DistilBERT} and fine-tuned it on the SemEval2017-STS dataset~\cite{nlp/Cer2017SemEval}. Both classifiers give a higher score for more equivalent sentences (e.g., \textit{The bird is bathing in the sink.} vs. \textit{Birdie is washing itself in the water basin.}) and a lower score if two sentences are dissimilar (e.g., \textit{A boat sails along the water.} vs. \textit{The man is playing the guitar}), all definitions and examples from \cite{nlp/Cer2017SemEval}.
\ci{We measured the correlation between the classifiers' semantic similarity estimation to human annotations as well as their response latency in comparing 100 pairs of sentences. The detailed implementation and evaluation procedure are given in Appendix~\ref{appendix:ssc}.}
% The similarity predictions of the classifier reflect human similarity judgments well achieving correlations $>80\%$ with manually labeled ground truth data. The details on training and evaluation are given in the Supplementary Material.

The results show that the similarity predictions of \ci{both classifiers} reflect human similarity judgments, well-achieving correlations $>80\%$ with manually labeled ground truth data. \ci{However, the fine-tuned classifier is faster, spending on average 2.64 seconds (SD = 0.28), while the prompt-based classifier spent on average 5.16 seconds (SD = 1.20). Accordingly, we conclude the fine-tuned classifier is more suited to our interaction scenario and used it for our chatbots.}

Internally, the fine-tuned classifier predicts a score on a $5$-point scale. Based on comparison with human annotation, we split this core into three ranges for our application when judging pairs of user-submitted ideas: for a similarity score below 2, the ideas are marked as dissimilar, above 2 as similar, and above 3 as too similar.

\subsection{Multi-armed bandit}

\begin{figure}[b]
  \centering
  \includegraphics[width=\linewidth]{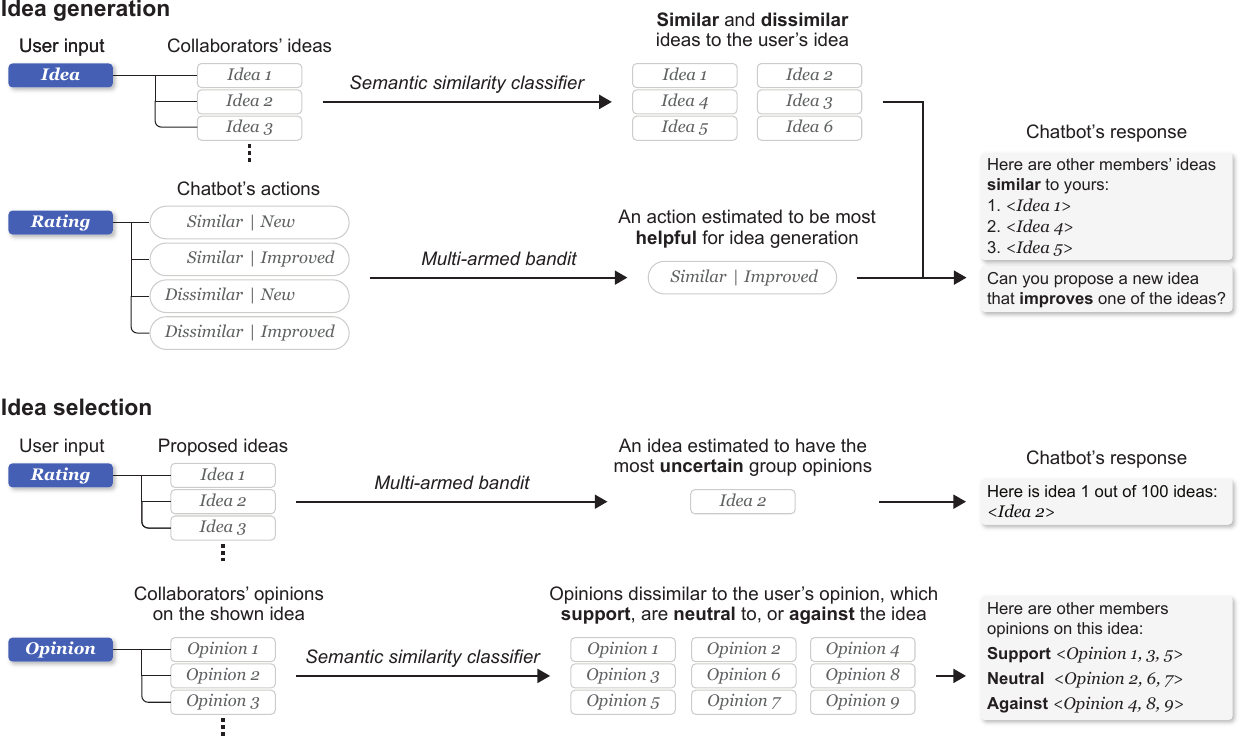}
  \caption{The system structure of the adaptive facilitator. During idea generation (top), the MAB system selects inspirations and ideation methods based on the individuals' rating of their own ideas. During idea selection (bottom), the MAB system presents ideas based on the users' collective rating of each idea (i.e., prioritizing ideas with uncertain group opinions).}
  \label{fig:system_structure}
\end{figure}

% MAB for the bandit-based facilitator
We designed the multi-armed bandit (MAB) system for the adaptive facilitator's behaviors. Its system structure is shown in Figure~\ref{fig:system_structure}.
The name originates from a context where a gambler visits a casino to play slot machines, so-called one-armed bandits. By playing one machine (action) at a time (trial), the gambler wins a certain amount of payout (reward). 
The gambler's goal is to win as much reward as possible within the finite number of trials by selecting the machine that gives the highest reward.
Since the gambler does not know which one gives more rewards, they need to try out different slot machines to find the most rewarding one (exploration) and mostly play the best one (exploitation), while balancing how much they should explore and exploit within the limited trials.

% Motivation: why MAB?
Our adaptive facilitator's objective aligns well with the MAB system. The facilitator tries one dialogue (action) at a time and receives a user reply (reward). Within the limited duration of the conversation, the facilitator needs to identify which dialogue is most helpful to the users (exploration) and perform the most helpful one (exploitation).

% \subsubsection{Overview of the algorithm}
% overview of the algorithm
% Which algorithm do we use?
We use the Upper Confidence Bound (UCB) algorithm~\cite{rl_book} for our MAB system.
At every trial, the algorithm computes two types of information for each action: 
% estimated reward and uncertainty. 
The estimated reward, representing \emph{how much reward the system would get by selecting the action in the next trial}, and the uncertainty, representing \emph{how much the rewards vary for an action across its trials}.
% This is computed based on all the rewards that each action has collected in the past.
% the MAB system will eventu
Based on the UCB formulation, even if the MAB system exploits the most rewarding action, it will eventually select another because of its high uncertainty.
% How does it compute the reward of each action?
A chatbot interaction guided by the MAB looks like this:
\begin{enumerate}
    % \item MAB selects an action ($A_t$) at trial ($t$);
    \item the MAB system selects one of the action \ci{(e.g., present similar ideas and request a new idea)};
    \item the chatbot performs the action;
    \item the user responds (e.g., generates and rates an idea);
    % \item MAB estimates the mean reward of the action ($Q_t(a)$);
    \item the chatbot receives a certain reward based on the quality of the user's response;
    \item the MAB system computes the estimated reward and the uncertainty of each action;
    \item the MAB system selects the next action that has the highest mean reward and uncertainty.
\end{enumerate}

We evaluated how our MAB systems would explore and exploit during each collaborative ideation phase in a simulated environment. We saw that the MAB chose the actions in a reasonable way, quickly identifying a suitable inspiration method (exploitation) but also changing the inspiration method if it became less beneficial (exploration), as if the user starts generating less helpful ideas from that type of inspiration. The details of our MAB system and its evaluation are given in Appendix~\ref{appendix:mab}.

\section{Empirical Evaluation}
\begin{figure}[b!]
  \centering
  \includegraphics[width=\linewidth]{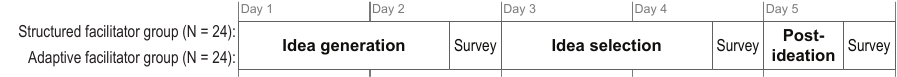}
  \caption{The structure of our user study. Two groups of participants performed idea generation and selection by interacting with the structured or adaptive facilitators. They reviewed the selected ideas on the last day. Participants shared their perceptions of chatbots' facilitation on a survey after each ideation phase.}
  \label{fig:study_structure}
\end{figure}

We conducted a user study to \ci{understand the strengths and weaknesses of our chatbots in facilitating asynchronous ideation.}
For this, we staged an asynchronous ideation event, where two groups of participants ideated with one of the chatbots at their own comfortable time and place (Figure~\ref{fig:study_structure}). We set the event goal as exploring ideas that can improve people's daily life during a pandemic and recruited participants who were interested in the subject.
% This resembles the early stage of co-design where the design subject is known, but the specific end-products and stakeholder groups are not yet well-defined~\cite{lucero2012, lee:codesign:2008}.

% This resembles the early stage of co-design where the design subject is known, but the specific end-products and stakeholder groups are not yet well-defined. \dis{The resulting ideas will set design directions for the later co-design stages.}

The ideation event consisted of three phases: idea generation, idea selection, and post-ideation. We \ci{held each idea generation and selection for two days to simulate the conventional asynchronous ideation held for multiple days, allowing participants to ideate at varying hours.}
% There was no direct communication between the participants, and everyone remained anonymous.
The post-ideation was for showing the results (three ideas with the highest ratings) to the participants and collecting their responses to the results. Throughout the event, participants interacted with chatbots only (i.e., our chatbots mediated the idea exchange among the participants).
% 
% At the end of each ideation phase, we collected participants' responses to the chatbots' facilitation using a survey with open-ended questions.

\subsection{Apparatus}
% Telegram messaging platform
We used the Telegram messaging platform to develop our chatbots (Figure~\ref{fig:apparatus}). Due to its popularity\footnote{\url{https://telegram.org/faq\#q-what-is-telegram-what-do-i-do-here}}, we could easily deploy our chatbots and avoid making participants learn a new interface. We implemented our chatbots using Python and connected to the Telegram platform through its API\footnote{\url{https://github.com/python-telegram-bot/}}. The participants could interact with the chatbots via the Telegram application on their PC or smartphone, using their personal accounts.

In addition to the regular message input field, we implemented response buttons for participants to easily request the chatbots to show other ideas, pause and resume the interaction, and rate ideas. This also enabled the chatbots to collect users' ratings on ideas between 1-7.
To identify the top three ideas in the end, we implemented a feature that computes a grand score of each idea based on its mean ratings and certainty (i.e., standard error of the mean ratings). In this approach, the chatbots avoided selecting ideas that only had a couple of maximum ratings.

\subsection{Participants}
We recruited 48 participants \ci{via social network} (Mean age = 27.21, SD = 4.95, self-identified as men = 21 and women = 27), who had experienced any type of collaborative ideation (e.g., group discussion). The participants varied in their backgrounds such as teachers, software developers, researchers, and nurses. We randomly assigned half of them to interact with either the structured or adaptive facilitator, considering gender balance. For spending minimum 100 minutes over five days, we compensated each participant with a 30-euro voucher. All participants received the same amount, unrelated to the number of ideas they generated or reviewed.

\subsection{Task}
For each day, participants chose when and for how long (minimum 20 minutes in total) to interact with their chatbot facilitators. They were free to pause and resume the interaction as they wanted. Following our chatbots' guidance, participants aimed to generate and evaluate as many ideas as possible. During the post-ideation, participants reviewed the three selected ideas and reported their satisfaction on a survey. At the end of each ideation phase, participants completed a survey to measure the usefulness of chatbots and comment on their facilitation.

\subsection{Measurements}
To understand the potential of chatbot facilitators, we observed 1) participants' interaction behavior with the chatbots, 2) performance in ideation, 3) perception of the chatbots' facilitators, and 4) satisfaction with the end result. 
For this, we recorded the moments when participants generated and selected ideas as well as the perceived quality of ideas.
% CSI and SUS
We measured participants' perceptions of the chatbots using the system usability scale (SUS)~\cite{sus} and creative support index (CSI)~\cite{csi}, which also measures the helpfulness of systems on collaborative activities.
% CSI measured the chatbots support in six themes related to collaborative activities: collaboration, enjoyment, exploration, expressiveness, immersion, and result worth effort.
% Open-ended
We collected participants' comments using open-ended questions, inquiring about what they liked and disliked throughout the asynchronous ideation.
% \revision{We clustered their comments to understand the trade-offs of chatbot facilitation compared to their prior experience of ideating with in-person human collaborators.}
% 
% Satisfaction
Lastly, we measured participants' satisfaction with the selected ideas with a survey inspired by the guidelines in consensus-building~\cite{susskind:book:1999,coleman:book:2014}. Even if participants believe that the selected ideas are not good enough, they can still be willing to commit to the group decisions. This is considered a successful group effort. Accordingly, we adopt the questionnaires tested in the previous study about consensus building \cite{shin:uist:2022}, which are answered with a 7-point Likert scale (Figure~\ref{fig:human_vs_chatbot}, bottom).

\subsection{Procedure}
We conducted the entire study online. 
% Participants individually began the interaction at their own time.
% 
Prior to the study, participants were given a manual for interacting with the chatbot facilitators and following the overall study schedule.
On each morning, research moderators reminded participants about the ideation via email. They sent another reminder an hour before the end of each ideation phase, informing participants to wrap up and complete the survey. All other ideation-related facilitation was led by the chatbot.
% They introduced the ideation goal, guidelines, and when each ideation phase ends.
%
During the post-ideation phase, the chatbots informed that three ideas with the highest ratings were selected and presented them with exemplary opinions from three categories (support, neutral, and against).
At the end of each phase, the chatbots conducted the surveys by explicitly informing participants to answer only based on their experience during the ideation, not considering the survey as a part of the interaction.
% 
% The study procedure was the same for both rule- and bandit-based facilitators.
% Is it?
% notable ideas = similar ideas + dissimilar ideas
% total ideas = notable ideas + too similar ideas.
% I tested my hypothesis and I think this is what I did. I actually compared each idea to all other generated ideas. If an idea has a semantic similarity score equal to and above 3 only to itself, then we keep the idea as a 'notable idea'. In contrast, If an idea's semantic similarity score is equal to or higher than 3 for the other ideas, then we filter it out. 

% change idea review phase as -> post-ideation phase.

% Example of very similar ideas
% 'Automatically sanitizing the door handles.', 'A public space that can automatically sanitize all visitors at the entrance.', '3.43'

% Example of similar ideas
% 'Automatically sanitizing the door handles.', 'Sanitized food containers for delivery.', '2.40'

% Example of dissimilar ideas
% 'Automatically sanitizing the door handles.', "Let's use more soft materials for the masks.", '0.65'

\section{Results}
\begin{figure}[b!]
  \centering
  \includegraphics[width=\linewidth]{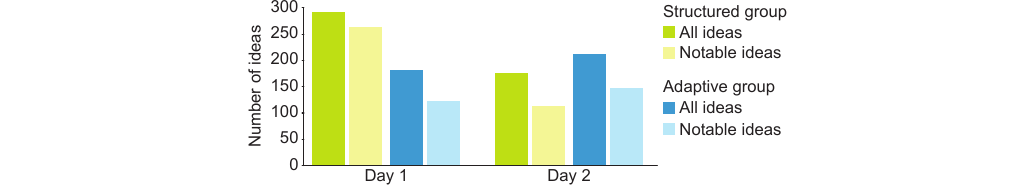}
  \caption{Number of all and notable ideas generated during the idea generation phase. Ideas that were not filtered out by our chatbots were considered notable.}
  \label{fig:numIdeas_ratings}
\end{figure}

% The general positive things to share at first.
The participants reported both chatbot facilitators to be helpful in generating and selecting ideas. Each group of participants generated 474 and 395 ideas by interacting with the structured facilitator (i.e., structured group) and the adaptive facilitator (i.e., adaptive group), respectively (Figure~\ref{fig:numIdeas_ratings}). Half of the participants from the structured group and nine participants from the adaptive group generated more than 20 ideas in total. The chatbots filtered out too similar, repetitive ideas and prompted the structured group to review 311 ideas and the adaptive group to review 200 ideas. Each participant reviewed between 21 and 68 ideas. 
% Two participants from the rule-based group were enthusiastic about the process and reviewed 109 and 231 ideas each.

The top three ideas from the structured group were \textit{``More widely available cheap solutions to grow food indoors", ``Some kind of online collaborative art project would cheer up people", and ``More communitarian activities within different parts of town to improve local services and advance a strengthened sense of togetherness". The adaptive group's ideas were \textit{``Development of more flexible, comfortable tools to test if we caught some type of flu would be better", ``Moisturizer that sanitizes your hands at the same time", and ``A simple website that has the rules and guidelines by officials that are effective now and a bot to answer people's questions".}}

\subsection{Interaction with the chatbot facilitators}
\begin{figure}[t]
  \centering
  \includegraphics[width=\linewidth]{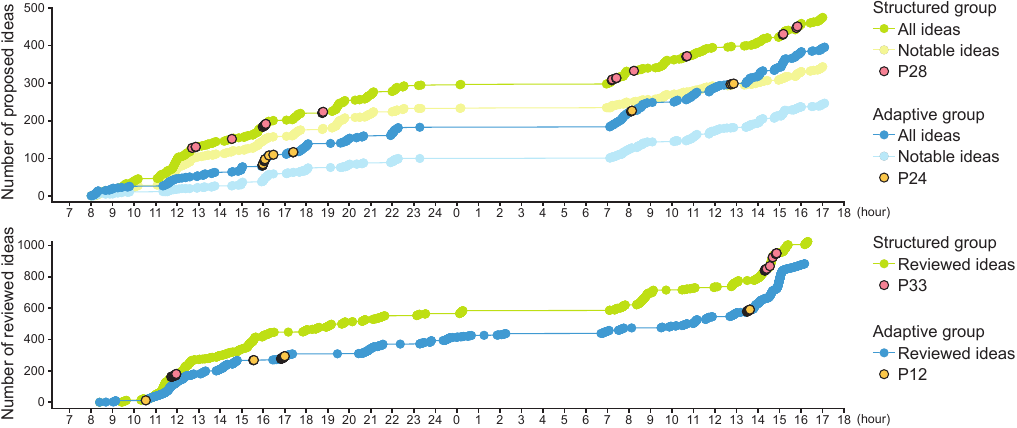}
  \caption{The moments that participants proposed ideas (top) and selected ideas (bottom). We highlight exemplary participants' interaction data. The data shows the participants' asynchronous collective effort spent at their own comfortable time.}
  \label{fig:interaction_timeline}
\end{figure}

To understand how participants collectively ideated over multiple days, we plotted their moments of interactions on a timeline (Figure~\ref{fig:interaction_timeline}). The data shows that both chatbot facilitators were interacting with some participants almost all the time. The individuals' data shows that they were indeed contributing their ideas on their own time, pausing and resuming their ideation. 
% example
For instance, P24 initiated the idea generation around 4 pm, shared ten ideas, took a break for 10 minutes, and shared three more ideas in 10-minute intervals. He came back to the chatbot after an hour to share four more ideas on the first day.
% In the second day, he started idea generation in the morning and shared his last three ideas after four hours. 
% 
The data also shows an increasing number of notable ideas proportional to the number of total ideas. Accordingly, we conclude that asynchronous ideation was achieved via the chatbot facilitators.

% We visualize the collective effort during idea selection by the number of opinions for each reviewed idea (Figure~\ref{fig:num_opinions}). The rule-based facilitator presented the ideas in the same order for all participants whereas the bandit-based facilitator adaptively presented the next ideas to review, making each participant review the ideas in a different order. In response, the rule-group collected more opinions across a fewer number of ideas (among 311 ideas, 68 ideas had more than 3 opinions) while the bandit-group collected fewer opinions across all 200 ideas (3 to 8 opinions per idea). Despite the differences, both groups of participants displayed above-average satisfaction with the selected ideas at the end (Figure~\ref{fig:satisfaction}).

% \begin{figure*}[h]
%   \centering
%   \includegraphics[width=\linewidth]{figures/num_opinions.png}
%   \caption{The number of opinions on each idea from the rule-group (left) and bandit-group (right). On the x-axis, the ideas are listed by the participants' self-rating of their own ideas. Ideas toward the left mean receiving a higher rating. The rule-group collected more opinions across fewer ideas while the bandit-group collected fewer opinions across all ideas.}
%   \Description{}
%   \label{fig:num_opinions}
% \end{figure*}

\subsection{Idea generation}
The goal of idea generation was creating as many ideas as possible by proposing new and improved ideas. To observe potential trade-offs between the chatbots' facilitation, we compared the results between the structured and adaptive group (between-group) and between the days (within-group).
% Say how we analyzed the data
For this, we performed two-by-two factorial analysis, performing Mixed ANOVA ($p$ $<$ 0.05) on IBM SPSS statistics software.

\subsubsection{Ideas similarity}
We analyzed the similarity of all ideas by comparing them against a single idea, which we added as one of the initial inspirations in this study (\textit{``A clear mask that allows people to see your expressions''}). The structured and adaptive group's mean idea similarity was 0.93 (SD = 0.4) and 1.41 (SD = 0.4), respectively (Figure~\ref{fig:idea_diversity}, left). 
According to our semantic similarity classifier, both groups mostly generated \emph{dissimilar} ideas (i.e., similarity score below 2).

% Chatbot facilitators
The statistical analysis shows that there was a significant main effect of chatbot facilitators on the similarity of ideas ($F$(1, 358) = 60.93, $p$ $<$ 0.01). This indicates that if we ignore on which day the ideas are generated, the structured group generated more dissimilar ideas. 
% Days
There was no significant main effect of days ($F$(1, 358) = 0.26, $p$ = 0.61)). 
% Interaction
Yet, there was a significant interaction between the types of chatbots and the days ($F$(1, 358) = 6.01, $p$ = 0.02). 
% This effect tells us that the types of chatbot facilitator had a different effect on the similarity of ideas that were generated in each day. 
The similarity distribution shows that the structured group generated less dissimilar ideas on the second day while the adaptive group generated more dissimilar ideas on the second day.

\subsubsection{Self-rating on the generated ideas}
\begin{figure}[t!]
  \centering
  \includegraphics[width=\linewidth]{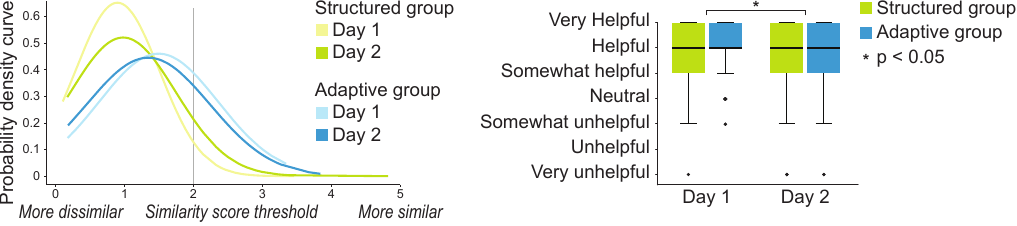}
  \caption{The similarity of ideas that participants generated by interacting with the structured and adaptive facilitators on the first and the second day (left) and the participants' self-rating of own ideas (right).}
  \label{fig:idea_diversity}
\end{figure}

% What is this measure?
To focus participants' attention on the subject, both chatbots asked the participants to rate the helpfulness of their ideas in achieving the ideation goal (Figure~\ref{fig:idea_diversity}, right).
% Rule vs AI
%% There was no statistically significant effect of chatbots.
The statistical analysis shows no significant main effect of chatbot facilitators ($F$(1, 358) = 1.93, $p$ = 0.17) or the interaction effect between the chatbot types and the days ($F$(1, 358) = 0.92, $p$ = 0.34)
% Day 1 vs Day 2
%% There was a statistically significant effect of days on the self-ratings. The ideas generated on the second day was considered as more helpful ideas.
Yet, there was a significant main effect of days on the self-ratings ($F$(1, 358) = 5.69, $p$ = 0.02), which indicates that both groups considered their ideas more helpful on the second day.
% Interaction
%% There was no statistically significant effect.
% There was no significant interaction between the types of chatbot facilitator and the days (F(1, 358) = 0.92, p = 0.34).

\subsection{User ratings}
\begin{figure}[t!]
  \centering
  \includegraphics[width=\linewidth]{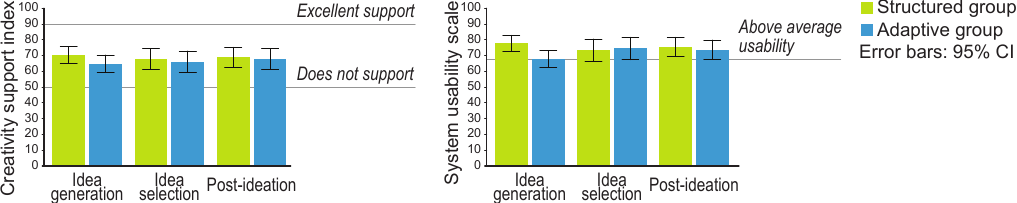}
  \caption{User response on the Creativity Support Index (left) and System Usability Scale (right).}
  \label{fig:csi_sus}
\end{figure}

The results of CSI and SUS are shown in Figure~\ref{fig:csi_sus}, grouped by the ideation phases. We assumed that the participants' perception of the chatbots could be changed based on their satisfaction with the end-result. Accordingly, we compared the user response between the chatbots (between-group) and between the ideation phases (within-group), performing two-by-three factorial analysis.

\subsubsection{Creativity Support Index}
The mean CSI score of the structured facilitator was 70.36 (SD = 2.67), 67.82 (SD = 3.28), and 68.86 (SD = 3.13) for idea generation, idea selection, and post-ideation, respectively. The mean CSI score of the adaptive facilitator was 64.80 (SD = 2.73), 66.13 (SD = 3.35), and 67.88 (SD = 3.20).
% no statistical significance
No statistically significant effects were observed across all comparisons, which indicates that participants' perception of chatbots did not change throughout the ideation or differed by the chatbots.
% There was no significant main effect of the chatbot facilitators (F(1, 45) = 0.68, p = 0.41) or the ideation phases (F(2, 90) = 0.17, p = 0.85). Likewise, there was no significant interaction between the chatbot facilitators and ideation phases (F(2, 90) = 0.52, p = 0.60). 
% conclusion
Cherry and Latulipe \cite{csi} state that the CSI score below 50 means that the system does not support creative work and above 90 means excellent support. Accordingly, we conclude that both chatbots well facilitated the participants' creativity in the collaborative ideation.

\subsubsection{System Usability Scale}
The mean SUS of the structured facilitator was 77.92 (SD = 2.56), 73.65 (SD = 3.44), and 75.73 (SD = 2.98) for each phase. The adaptive facilitator's mean SUS was 68.04 (SD = 2.61), 74.67 (SD = 3.51), and 73.91 (SD = 3.04).
% no statistical significance
The same as the CSI result, there was no statistical significant effects across all comparisons.
% main effect of the chatbot facilitators (F(1, 45) = 1.38, p = 0.25) or the ideation phases (F(2, 90) = 0.25, p = 0.78). There was no significant interaction between the two independent variables (F(2, 90) = 2.27, p = 0.11).
% conclusion
Brooke~\cite{sus} states that the SUS score above 68 means above-average usability. Therefore, we conclude that both chatbots were considerably usable for asynchronous ideation.

\subsubsection{Result Satisfaction}
The participants' satisfaction with the selected ideas is shown in Figure~\ref{fig:human_vs_chatbot}, bottom. Both groups expressed above-average satisfaction. Observing individuals' responses revealed that six and five out of 24 participants from the structured and adaptive group disliked the selected ideas.
% Three participants from each group neither liked or disliked the selected ideas.
% No statistical significance
We performed one-way ANOVA to compare the user response between the groups. There was no significant effect of the chatbots in all survey items.
% (p > 0.1).
% 1 (F(1, 45) = 0.17, p = 0.68)
% 2 (F(1, 45) = 1.79, p = 0.19)
% 3 (F(1, 45) = 2.84, p = 0.10)
% 4 (F(1, 45) = 0.23, p = 0.63)

% In more details...
% By looking at the individuals' response, We found that six and five participants from each rule- and bandit-group disliked the selected ideas. Three participants from each group neither liked or disliked the selected ideas.
% We conducted a formal analysis between the groups and found a significant effect of the chatbot on their willingness to commit to the end result (F(1, 16) = 7.78, p = 0.14). 
% Despite their preference to the ideas, two participants expressed that they were willing to commit to the selected ideas.
% conclusion
% Whereas the survey data did not highlight the differences between the chatbot facilitators, the participants' comments on the selected ideas remarked the potential benefits and limitations of the chatbots, especially related to their different facilitation during the idea selection. We discuss this in the next section.

% \begin{figure*}[t]
%   \centering
%   \includegraphics[width=\linewidth]{figures/satisfaction.png}
%   \caption{User response to the selected ideas as the result of idea generation and selection.}
%   \Description{}
%   \label{fig:satisfaction}
% \end{figure*}

\subsection{Experience of chatbot facilitators}
% Remind about our design
% Both chatbots (i) showed other group members' input as inspiration, (ii) suggested ideation method, and (iii) requested the rating on the ideas.
We analyzed the participants' comments via affinity diagraming~\cite{lucero:interact:2015}.
Despite the differences between the ideation phases and the chatbots, both groups of participants shared similar positive and negative comments. Below, we grouped them by our chatbot designs.

\subsubsection{Inspiration}
The participants mostly commented on the inspirations brought by the chatbots.
% Positive
They considered reading others' ideas and opinions 
helped them diversify ideas, propose more creative ideas, and more correctly evaluate ideas.
% examples for IG
For instance, P27 commented, \textit{``It showed ideas from other collaborators and spiked my creativity.''}
Likewise, P48 commented that the dissimilar ideas inspired new ideas that he would not have considered.
% The participants also commented that seeing other group members' opinions inspired them to re-evaluate the ideas from others' perspectives. 
Based on such responses, we confirm that the chatbots facilitated participants to build on each other's contributions in the asynchronous setting.

% Negative 
Yet, negative aspects were aslo reported that building on the others' ideas was difficult when they were written unclearly. P34 commented, \textit{``Sometimes the participants didn't express or describe their ideas properly. If there was someone modifying the errors, then show to other participants, that would be better.''} Other participants (P2 and 25) reported that some opinions were out of context (e.g., \textit{``I agree with the second opinion.''}) and stunted their own thought processes.

% % showing the ideas by their similarity.
% \revision{The participants also commented on judging ideas as similar, which the facilitators had presented as dissimilar. Nevertheless, the participants found both ideas useful.} P2 reported that they encouraged refining her own ideas, and P48 commented that the dissimilar ideas inspired him about new ideas that he would not have considered.
% %
% Interestingly, one participant (P42) from the rule-group reported seeing only the dissimilar ideas was demotivating (the first day). He commented, \textit{``I felt that my ideas weren't heard, the other ideas were nothing like my ideas.''}

\subsubsection{Ideation method}
% Suggesting the ideation methods was rarely recognized by the participants.
% Positive
The participants reported that the ideation method made them think more productively with a specific aim. For instance, P36 commented that the chatbot's suggestion guided her to improve the ideas rather than quickly criticizing them. Likewise, P21 commented, \textit{``I particularly liked how I could give my opinion on an idea before hearing others opinion... I could correct myself with the opinions from the group.''} No negative responses were reported about the chatbots suggesting ideation method.
% P28 commented that she appreciated that the chatbot suggested perspective-taking and asked her about any changes in her opinions after seeing the others' opinions.

% Does it mean that suggesting ideation method was not impactful?
% Despite the small number of comments, we could observe the impact of suggesting ideation method from the diversity of generated ideas. We reflect on this in the later subsection.

% \subsubsection{Rating}
% The participants shared no comments about the chatbots asking them to rate their ideas.
% % 
% This chatbot behavior was designed to prompt users about the design goal.
% % Reflection about why there were no comments on this.
% We assume that the reason for no comments on this chatbot behavior might have been due to people's familiarity of voting approach in making group decisions, hence not worth mentioning compared to the chatbots' other two behaviors. 
% % Another assumption is that the other group members' ideas might have already influenced participants to be constantly aware of the design goal, hence sensing less impact from the rating. 
% We conclude that further investigation is required to understand the how rating procedure can rtain users' attention on the design goal.

\subsection{Experience of chatbot-facilitated asynchronous ideation}
% What is this?
The analysis also revealed the strengths and weaknesses of asynchronous ideation facilitated by the chatbots. Below, we present the five themes.

\subsubsection{Anonymity}
% Positives
The participants appreciated that the collaborative ideation was performed anonymously. They reported that being unaware of who the others were encouraged them to share even less practical but more creative ideas, without worrying about the others' criticism. The most representative response was from P28, \textit{``I liked that the ideation was anonymous so it was easier to suggest even slightly odd ideas.''} The anonymity was also considered helpful during idea selection. The participants reported that they could evaluate the ideas with more objective viewpoints, instead of hiding their true opinions to not offend the others. For instance, P20 commented that collaboration without being scared of hurting someone's feelings was helpful. 
After seeing the selected ideas, P1 remarked, \textit{``This tool was very democratic in showing ideas, which helped developing ideas without being biased by whom it came from."}
% There were no negative comments about the anonymous collaborative ideation.

%P26: I liked that the ideas were anonymous so it was easier to suggest even slightly odd ideas. I also liked that I did not have to come up with the ideas at one meeting

\subsubsection{Asynchrony}
% Positives
For all participants, this was their first time attending asynchronous ideation. 
In response, the participants acknowledged the benefit of asynchronous settings and the chatbot faciliation. They reported that the chatbots enabled contribution at their own comfortable hours and pace. P19 commented, \textit{``It was easy to share ideas, it gave me a chance to work and collaborate with my preferred times and rhythm.''} The others also remarked that they could spend as many hours as they needed for generating ideas and appreciated that they did not have to come up with all ideas in a single meeting.
%
% The participants' interaction data (Figure~\ref{}) also support these responses.

%P19: It was easy to share ideas, it gave me a chance to work and collaborate with my preferred times and rhythm. Seeing others' ideas gave me ideas

% Negatives
However, the participants reported that the lack of immediate feedback lowered the immersiveness of collaborative ideation. P40 commented that he would have enjoyed the idea generation more if there was profound connection with the other group members. Similarly, P47 expressed the urge for meeting the people who generated the ideas that he liked. 
Another limitation was the unequal amount of available inspiration between the participants. In the beginning of each ideation phase, there were relatively fewer ideas and opinions. In response, 11 participants reported seeing the same inspirations repetitively, which demotivated their ideation.

\subsubsection{Accessibility}
% With what? Where did they interaction?
Most participants interacted with the chatbots via their smartphones, except for seven participants who used laptops. In response, the participants reported that they ideated in diverse places such as a couch, bed, desk, park, and public transportation systems.
% Postive
Since the participants could interact with the chatbots at any place and time, they acknowledged the chatbots' readiness for collecting ideas and inspiring their ideation. P25 and P4 commented that they could swiftly interact with other group members' opinions. P1 commented, \textit{``I liked that the bot was constantly present. I did not have to wait long to see other’s ideas.''} Likewise, P27 reported that being able to track his ideas whenever he needed was beneficial to resume the ideation.

% Negative
% None. However, because the chatbots enalbed swift way of interaction a few participants displayed a concern that other participants might participate in the collaborative ideation sincerely.
% However, we noticed that the readiness can potentially pose a wrong impression about the chatbot interaction, that the collaborative ideation is nothing serious.

\subsubsection{Text only ideation}
Mostly negative responses were reported about the text-only ideation. The participants commented difficulties in describing and developing ideas without any other modalities such as drawings or video. P2 commented, \textit{``I like brainstorming visually, so writing purely by text was a little stiff at times.''} Likewise, P25 and 9 commented that understanding others' ideas and thought processes only based on their writings was difficult.

\subsubsection{Non-human facilitator}
The participants highlighted the benefits of talking to the chatbots, instead of talking to the other group members or human faciliator. For instance, they appreciated having no social pressure in making the `listener' wait and the chatbots' polite responses regardless of what they contribute. At the same time, the participants shared the limitations of being a chatbot such as the lack of expression, repetitive dialogue, and the shallow conversation. These attributes were considered to make the interaction boring, lowering their commitment to the ideation. P35 commented that his perception of interacting with the chatbot made him less serious during the ideation.

\section{Expert Evaluation}
% Study 1: Can chatbots be really useful from the co-design participants' perspective?
% Study 2: Is there anything that our chatbots did good, bad, or missed from human facilitators' standards? Because experts know what could have been done better, 
% We ran an expert evaluation study to explore the potential of chatbot facilitators from human facilitators' perspectives.
\ci{The empirical evaluation study} revealed the pros and cons of the chatbots' facilitation from the collaborators' perspective.
To extend the findings, we conducted an expert evaluation study to explore chatbots' potential from human facilitators' viewpoints. For this, we recruited an expert with a unique profile who has studied and facilitated collaborative ideation for over 25 years (age = 49, self-identified as a woman). In particular, the expert has experience in facilitating collaborative ideation events with hundreds of participants both in in-person and asynchronous settings, which can provide critical insights about human facilitators' challenges and the design of chatbot facilitators.

To prime the expert for reviewing the chatbots' facilitation, we had the expert first facilitate an ideation workshop independently, without using chatbots. The workshop was held for an hour with 11 participants (Mean age = 27.18, SD = 7.55, self-identified as men = 8 and women = 4), following the same ideation structure and goal as our chatbots facilitated. Afterward, we demonstrated our chatbots and conducted an interview to collect the expert's perspectives.
% We also recorded the expert's behaviors throughout the workshop.
\ci{We also recorded the participants' ideas and their satisfaction with them using the same survey in the earlier study. We used them as a baseline to further analyze the chatbot-facilitated ideation.}

\begin{figure}[t!]
  \centering
  \includegraphics[width=\linewidth]{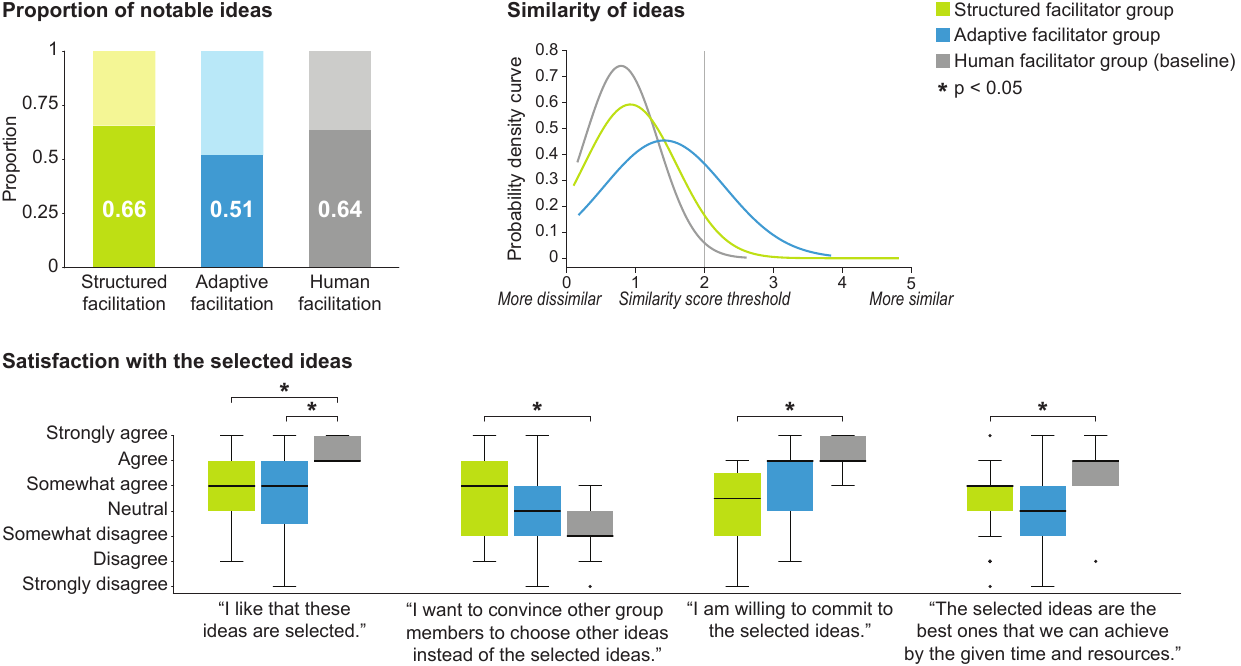}
  \caption{The comparisons between the human-facilitated and chatbot-facilitated ideation.}
  \label{fig:human_vs_chatbot}
\end{figure}

\ci{The comparisons between the chatbot- and human-facilitated groups are shown in Figure~\ref{fig:human_vs_chatbot}.
We compared the \emph{proportion of notable ideas}
% , instead of comparing the total number of ideas, as human-facilitated ideation is held relatively shorter than the chatbot-facilitated ones. We compared each pair of conditions 
using the chi-square test of independence with a Bonferroni correction. We compare the groups' \emph{idea similarity} and \emph{satisfaction with the selected ideas} using the Kruskal-Walis H test with Dunn's test as a post-hoc analysis. Below is the summary of the results:
\begin{itemize}
    \item Both chatbot-facilitated groups' proportion of notable ideas was statistically similar to the human-facilitated group's (both $p$ $>$ 0.1). 
    \item The structured group's idea similarity was statistically similar to the human-facilitated group's ($p$ = 0.80), while the adaptive group's idea similarity was statistically higher, indicating less idea diversification ($p$ $<$ 0.01).
    % , while the adaptive group's idea similarity was statistically higher, which indicates that the adaptive group's ideas were more similar to each other ($p$ $<$ 0.01).
    \item The adaptive group's satisfaction with selected ideas was statistically similar to the human-facilitated group's, except for their preference for the selected ideas. Both chatbot-facilitated groups had statistically lower preferences than the human-facilitated group (both $p$ $<$ 0.01).
\end{itemize}
The results suggest that the structured and adaptive facilitator has their own strengths, achieving idea diversification and satisfaction similar to human-facilitated ideation, respectively.}

% Whereas there was a statistically significant difference between the chatbot-facilitated group ($\chi^2$(1) = 10.30, $p$ $<$ 0.01), both groups generated a statistically similar proportion of notable ideas to the human-facilitated group.
% Idea similarity
% We compare the groups' idea similarity and satisfaction with the selected ideas using the Kruskal-Walis H test with Dunn's test as a \emph{post-hoc} analysis.
% as the human-facilitated group's data failed the Shapiro-Wilk normality test ($p$ $<$ 0.01). 
% The results showed that the idea similarity between the chatbot- and human-facilitated groups was statistically different ($H$(2) = 77.73, $p$ $<$ 0.01). In particular, the human-facilitated group generated more dissimilar ideas than the adaptive group ($p$ $<$ 0.01).
% Satisfaction

\ci{The interview with the expert highlighted pain points in facilitating multiple groups of collaborators. The expert described that human facilitators get tired throughout facilitation, making it challenging to pay closer attention to individuals. The expert added that she also did not have enough energy to take care of less active participants during this workshop. Yet, hiring more trained facilitators is expensive.}

Accordingly, the expert remarked on the potential hybrid facilitation between human and chatbot facilitators that complement each other's limitations. She remarked on the capability of chatbot facilitators for (i) cross-pollinating ideas, (ii) supporting ideation with facts, and (iii) moderating progress. While ideating with the participants, the expert often talked about ideas from the other group as inspirations. She commented that this process could have been supported with chatbots. She said, \textit{``I tried to cross-pollinate ideas by bringing ideas between groups, but I could not do that systematically because I cannot hold all of them in my head. This is something that the chatbots can do.''}
In addition, she commented that there were moments when knowing more facts about certain ideas could have helped the participant improve ideas, which chatbots could proactively provide. We also observed that the participants often searched for images on their smartphones to enrich their discussion.
% She said, \textit{``A chatbot could help with facts. If an idea has been mentioned, the chatbot could say there is research that supports this.''}
% 
Lastly, she expected the chatbot to moderate the progress of ideation with prompts. She commented that chatbots could help teams understand where they are in the ideation process and keep them on track, preventing participants from just focusing on their own ideas.

The expert also highlighted the shortcomings of chatbot facilitators including (a) the lack of embodied interaction, (b) experience sharing, and (c) multi-modality. First, the expert remarked that facilitation is a performance and collaboration does not only happen at the idea level. She added that social interactions during collaborative ideation demand the whole body and spatial aspects that chatbots cannot replace.
Second, the human facilitator commented that anonymous collaboration through chatbots could demotivate participants. She remarked that people value meeting others and sharing experiences, which strongly motivates them to participate in collaborative ideation events.
Lastly, the human facilitator pointed out text-only ideation. She commented that ideation using only text is feasible in the early stage, but it would limit the development of concrete subjects or prototypes.

\section{Discussion}
Developing conversational agents that adopt human behaviors has become more accessible thanks to LLMs~\cite{roleplay/park2023generative, roleplay/Shanahan2023}. Still, this requires an understanding of what could be adopted from humans, instructing (e.g., prompting) behaviors that make the best out of conversational agents.
In this paper, we make a case for facilitating \ci{asynchronous} ideation by leveraging conversational agents.
% 
% Summary of our studies and what we learned
We presented two designs of chatbot facilitators based on the literature on human facilitators. 
We investigated their potential from the viewpoints of collaborators (Study 1) and an expert facilitator (Study 2), as shown in Table~\ref{tab:summary}. Study 1 showed that both chatbots enable asynchronous ideation among collaborators, allowing them to build on each other's 
contributions at their own pace.
Study 2 suggests that the structured facilitator has an advantage in diversifying ideas and the adaptive facilitator can help collaborators choose satisfying ideas similar to human-facilitated ideation.
However, the two studies also exposed a shortcoming of chatbot facilitators: they do not moderate social interaction among participants more broadly. In the following, we elaborate on this finding and reflect on its implications.

\begin{table}[h]
    \centering
    \includegraphics[width=\linewidth]{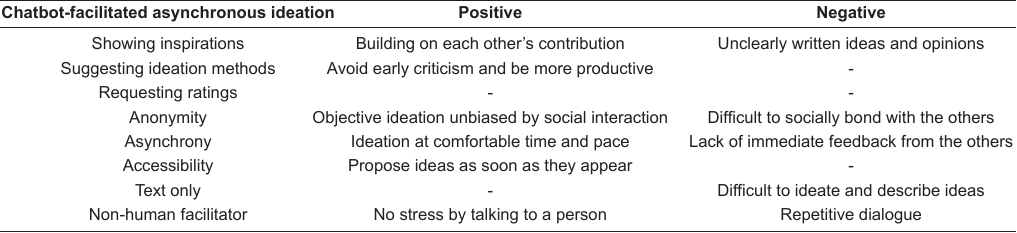}
    \caption{Summary of our findings on the positives and negatives of asynchronous ideation facilitated by chatbots.}
    \label{tab:summary}
\end{table}

% \begin{table}[h]
%     \centering
%     \small
%     \begin{tabularx}{\textwidth}{ccc{>{\centering\arraybackslash}c}} 
%         \hline
%         \textbf{Chatbot-facilitated collaborative ideation} & \textbf{Positive} & \textbf{Negative} \\ 
%         \hline
%         Showing inspirations & Building on each other's contribution & Unclearly written ideas and opinions \\ 
%         Suggesting ideation methods & Avoid early criticism and be more productive & - \\
%         Requesting ratings & - & - \\
%         Anonymity & Objective ideation unbiased by social interaction & - \\
%         Asynchronicity & Ideation at comfortable time and pace & Lack of immediated feedback from the others \\
%         Accessibility & Propose ideas as soon as they appear & - \\
%         Text only & - & Difficult to ideate and describe ideas \\
%         Non-human facilitator & No stress by talking to a person & Repetitive dialogue \\
%         \hline
%     \end{tabularx}
%     \caption{Summary of the positives and negatives of chatbot-facilitated collaborative ideation.}
% \label{tab:summary}
% \end{table}

% What our results mean for a co-designer?
\subsection{Applying chatbots in collaborative activities}
Our studies suggest that chatbots can facilitate collaborative ideation in asynchronous settings. We propose three practical suggestions on how to deploy chatbot facilitators to \ci{benefit collaborative efforts in creative projects such as design.}
First, we suggest using \emph{chatbots for aligning collaboration efforts}. In collaborative activities, it is essential that group members aim for the same ideation goal. However, human facilitators can experience challenges in closely engaging with individuals and redirecting their attention. Instead, as we observed in Study 1, chatbots can engage with all members to focus on joint goals and get them to reflect on their own contributions as a group.
% In Study 2, the human facilitator did this personally by reminding participants about the co-design subject. Chatbots can facilitate this. 
% 
%Our first study showed that chatbots can engage with individual participants and focus their attention on the co-design subject. Our chatbot facilitators made participants evaluate the helpfulness of their own ideas for satisfying the design goal. In response, all participants proposed ideas that are related to the co-design subject.
 
Second, we suggest using \emph{chatbots for guiding ideation toward unexplored subjects}. To diversify ideas, collaborators need to know what kinds of ideas they have created as a group. However, reviewing all generated ideas and identifying unexplored subjects is laborious even for multiple human facilitators. Instead, chatbots with advanced NLP methods could cluster ideas and identify less discussed subjects. \ci{Then, similar to our chatbots suggesting inspirations, chatbots could prioritize overlooked subjects to promote effective idea exploration.}
% Our chatbot facilitators filtered out similar ideas and presented notable ideas using semantic similarity classification. 
%For instance, semantic clustering could group similar ideas, which chatbots then identify smaller groups of ideas and encourage participants to further investigate them.
 
Third and last, we suggest using \emph{chatbots for moderating collaborators understanding of each other}. Study 2 showed that collaborative ideation is not just generating or selecting ideas together. We observed that the participants actively shared their personal experiences to empathize with each other and convey their perspectives.
% In co-design projects without chatbots, this is difficult since participants cannot interact with everyone.
% 
%Similar to collecting participants' ideas, we believe chatbots could also effectively organize and share participants' stories. Especially, 
Our study showed that anonymous idea exchange through chatbots is helpful in sharing personal thoughts without worrying about criticism.
Previous studies also show that users are more willing to share their personal stories with chatbots than other human beings, partially because chatbots appear less judgmental~\cite{lee:chi:2020,collins:dis:2022}. 
This could help collaborators share their thoughts more directly and understand each other more correctly.

% Fourth and last, we suggest using \emph{chatbots for building common ground among collaborators}.
% %Participants need to have a shared understanding of their progress so that they can also collaborate in the later stages of co-design. 
% %To this end, human facilitators often guide participants to reflect on their collaborative ideation at the end. 
% % 
% The two studies showed that chatbots can facilitate this process. Chatbots can track and compare individuals' ideas. Chatbots can collect participants' understanding of, for instance, a design direction and examine whether participants' understanding aligns with each other. 
% % Potentially, chatbots could also lead consensus-building in the case of conflicts among participants~\cite{kim:cscw:2021, shin:uist:2022}.

% Better title
\subsection{Understanding differences of chatbot facilitators}

Our studies provide directions for adapting human facilitators' behaviors to chatbot facilitators.
% choosing a suitable computational approach to a project. 
%The first study showed that the rule- and bandit-group generated different number of and diversity of ideas. However, there was no speicifc user response that could describe the differences. Nevertheless, we discuss our assumptions and suggest future studies.
%
% Why rule-group generated more diverse ideas?
We learned that chatbots can lead structured facilitation to the generation of more diverse ideas.
%This result might be related to the burden caused by considering multiple aspects of ideas. 
In particular, the structured facilitator guided the participants to think about new ideas on the first day only. In response, the participants could have focused on diversification without the need to consider their quality. In contrast, the adaptive facilitator suggested both diversifying and improving ideas in a single day. This might have encouraged the participants to consider both the dissimilarity and quality of ideas, overburdening their ideation process. This assumption aligns well with guidelines of effective brainstorming, having a single purpose of ideation at a time and suspending criticism~\cite{puccio:jcb:2020,paulus:fp:2018,thoring:desire:2011}.
\ci{We also learned that chatbots can facilitate adaptively, achieving a similar level of satisfaction with final ideas as human-facilitated ideation. We assume that guiding collaborators to review ideas with uncertain group opinions helped them strengthen their consensus. Potentially, combining both structured and adaptive facilitation into a single chatbot could be more effective, guiding one ideation method per idea generation and adaptively prioritizing ideas to review during idea selection.}

\subsection{Toward hybrid facilitation}
Based on our results, we suggest directions toward a `hybrid model' of human--AI facilitation that exploits their unique strengths.
First, we note that chatbot facilitators can be in a secondary role (an assistant) to a human facilitator. 
In Study 2, the human facilitator reported the moments when she could have applied an extra facilitator. 
She reported that moderating more than one group of participants was challenging and she often forgot to encourage less active participants.
% , and she wanted to check the facts related to proposed ideas.
% 
We expect chatbots to provide such support. For instance, while a human facilitator engages with one group of participants, chatbots could mimic the human facilitator's dialogue and guide the rest of the participants. 
Moreover, it is vital for the human facilitator to retain leadership and accountability. 
%Chatbots could also remind certain facilitator roles to human facilitators, similar to how our chatbots prompted participants about ideation rules. 
% 
Second, the techniques presented for chatbot facilitation so far in the literature suggest that chatbots are best when deployed in collaborative ideation. 
%Our studies showed that chatbots can guide a large number of participants. 
% 
%While chatbots conduct co-design activities, 
Chatbots can help human facilitators to monitor the progress and learn from it. Human facilitators could also use this period to focus on designing the next stage of collaborative ideation, identifying opportunities to include collaborators with different perspectives or exploring other areas of the design space~\cite{steen:ijd:2011}. 
%This could help participants effectively progress through multiple co-design stages.
% 
Lastly, chatbot facilitators could help human facilitators better understand the participants after the first round.
%Study 2 showed that one critical challenge for chatbot facilitators is social interaction. However, we expect that socially bonding with a large number of stakeholders would also be challenging for human facilitators. 
% 
Chatbot facilitators can identify the characteristics or preferences of participants for human facilitators by continuously engaging with each individual collaborator throughout collaborative ideation activities.
%Chatbots can provide such information to human facilitators before or during their interaction with participants so that human facilitators can plan how they engage with participants.

\subsection{Limitations and future work}
We report four limitations in our study and propose corresponding future work.
\ci{First, we focused on facilitating asynchronous ideation in the context of idea generation and selection. Whereas diversifying and converging ideas are the fundamental activities in the ideation process, how chatbots can facilitate other collaborative scenarios, such as problem-solving and deliberate discussion, in asynchronous settings remains unexplored. Future work could look into the different facilitation requirements that arise in different scenarios and examine how chatbots could adopt them.}
Second, we studied chatbots that used text as the medium. In Study 1, four participants reported that ideating and describing ideas using only text was challenging. In the future, chatbots could facilitate multi-modal communication. Chatbots are already capable of integrating sketches, images, videos, and audio into dialogue. Recent advances in multi-modal machine learning could provide the basis for chatbot facilitators to understand users' non-text contributions~\cite{clip}.
Third, our expert evaluation study is based on a single case study. While we recruited the expert with unique profile, we acknowledge that our expert's viewpoints may not fully generalize to other facilitators. To address this, future work could include experts with diverse backgrounds and competencies to uncover the broader perspectives on using chatbots as facilitators.
Last, we studied asynchronous and anonymous ideation. Despite our findings on the benefits of anonymity, having no direct communication and not knowing each other could demotivate collaborators. Future work needs to take this into account and study artificial facilitators that can also promote social interaction among collaborators.
\section{Conclusion}
\ci{In this paper, we designed chatbot facilitators to guide asynchronous idea generation and selection among collaborators. We designed the structured and adaptive facilitators by adopting the guidelines found in the literature about human facilitators. Our structured facilitator guided individuals by providing a structured ideation process and our adaptive facilitator guided them by adapting to their ideation performance.
Our studies suggest the strengths and limitations of chatbot facilitators. Both chatbots were found to be helpful, especially when helping collaborators build on each other's contributions. Each structured and adaptive facilitator had their strengths in diversifying ideas and achieving satisfaction with selected ideas similar to human-facilitated ideation. With our findings, we discuss the potential of a `hybrid model' where human and chatbot facilitators complement each other's strengths. We conclude that chatbots can be promising alternative facilitators of asynchronous ideation, providing continuous guidance in the absence of human facilitators.}

\begin{acks}
This research was supported by the Research Council of Finland grant 357578 (Subjective Functions); the Basic Science Research Program through the National Research Foundation of Korea (NRF) funded by the Ministry of Education (2022R1A6A3A03056886); and IFI program of the German Academic Exchange Service (DAAD).
\end{acks}

\bibliographystyle{ACM-Reference-Format}
\bibliography{coDesignAI}

\raggedbottom

\pagebreak

\appendix
\section*{APPENDIX}
\section{Semantic similarity classifier}
\label{appendix:ssc}

Using a semantic similarity classifier allows us to compare two user-provided ideas and automatically judge how similar or dissimilar they are. This removes the need for a human facilitator or for the participants to manually check the similarity of ideas and categorize them. Given the speed of the semantic similarity classifier, it also provides the chatbot with near-instant information on the similarity, avoiding long wait times and allowing for a more fluent conversation with the user. For this, we tested a \textbf{prompt-based classifier} based on LLM and a \textbf{fine-tuned classifier} based on a language model.

We implemented the prompt-based classifier using GPT-4 (gpt-4-0613), the latest model at the time of testing, and followed the prompt design introduced by Getto et al.~\cite{gatto:2023:stsGPT}. The authors evaluated the competence of two LLMs (Llama2-7b and gpt-3.5-turbo-0301) on a semantic textual similarity prediction using different prompt designs. Among them, we followed the best-performing design (\textbf{Prompt}: \textit{Output a number between 0 and 1 describing the semantic similarity between the following two sentences: Sentence 1: $<Text 1>$ Sentence 2: $<Text 2>$}). 

We implemented the fine-tuned classifier by fine-tuning DistilBERT~\cite{nlp/Sanh19DistilBERT}. We chose DistilBERT as it is a compressed version of the popular BERT~\cite{nlp/Devlin19BERT} model and, thus, reduces hardware requirements. We take the ``base-uncased'' variant and train it on the English portion of the SemEval2017-STS dataset~\cite{nlp/Cer2017SemEval}. We train for $5$ epochs with a batch size of $32$, fp16 precision, and mean-squared-error loss, selecting the weights of the best epoch according to the development set.

\subsection{Evaluation of semantic similarity estimation}
We evaluated the classifiers by measuring the correlation of their predictions with the manually annotated development and test set of SemEval2017-STS.
Fine-tuned classifier achieved a Pearson/Spearman correlation of $87/87\%$ on the development set and of $82/81\%$ on the test set. Prompt-based classifier achieved $84/85\%$ and $80/81\%$, respectively. While both not beating the state-of-the-art in NLP, this is in range with modern classifiers. Sanh et al., \cite{nlp/Sanh19DistilBERT} report, e.g., a correlation of $91\%$ on the development set. Accordingly, we conclude that both classifiers can provide high-quality semantic similarity estimation. 

\subsection{Evaluation of response latency}
We evaluated our classifiers' response latency by measuring how long it takes them to estimate the similarity between an input sentence and another 100 sentences. This resembles our chatbots' facilitation scenario, where a user submits an idea and the chatbots present the three most similar or dissimilar ideas from all other collaborators. The fine-tuned classifier performs the 100 comparisons in a single run, outputting a set of similarity scores.
In the case of the prompt-based classifier, we adjusted the prompt to instruct LLM to fetch the three most similar and dissimilar sentences from 100 sentences in relation to an input sentence (Prompt: \textit{Here is a set of sentences: Sentences: $<100 sentences>$ Select three most similar and dissimilar sentences to the sentence below: $<1 sentence>$}). We made this adjustment as Gatto et al.'s prompt design can not perform multiple comparisons in a single prompt~\cite{gatto:2023:stsGPT} (i.e., running prompts multiple times takes longer).

We tested each classifier 100 times and computed their mean response latency. The results showed that the fine-tuned classifier is faster (Mean = 2.64 seconds, SD = 0.28) than the prompt-based classifier (Mean = 5.16 seconds, SD = 1.20). Accordingly, we conclude that the fine-tuned classifier is more suitable for our interaction scenario.

\subsection{Estimating the threshold of the semantic similarity classifier}
\label{appendix:ssc_threshold}
We evaluated how well the fine-tuned classifier works in our idea-generation context and mapped its similarity scale to a categorization that is useful for us (i.e., similar or dissimilar with regard to user-provided ideas). For this;
\begin{enumerate}
    \item We generated 100 ideas written in colloquial styles that represent the result of our chatbot interaction (e.g. \emph{``I think more comfortable masks need to be designed''}).
    \item We obtained similarity scores for all sentence pairs from our classifier.
    \item Since the distribution of scores was skewed towards many dissimilar pairs, We randomly sampled 12 pairs from each 0.5 interval between 0 to 3.5 (7 range, 84 pairs in total): We excluded the range above 3.5 as such a score was only reached when the sentences were the same.
    \item 10 participants (Mean age = 24.80, SD = 4.26, 5 males) reviewed the selected pairs in a randomized order and rated their similarity using a 5-point Likert-scale (1 = very dissimilar and 5 = very similar): We excluded the pairs that were labeled as ``neutral'' and grouped together ``very similar'' and ``similar'' into a single ``similar'' label (and respectively for ``dissimilar'') since the distinction was not relevant for our use case.
    \item Based on their ratings, we thresholded the classifier's output score. We classified everything below and equal to the threshold as dissimilar and above as similar. We computed agreements between classifier and human labels via accuracy, i.e., the number of annotations where the classifier and human agree / the total number of annotations.
    \item The highest agreement to `similar' is reached for a threshold of $2.0$ with an accuracy of $83\%$. The highest agreement to `very similar' is reached for $3.0$ with $84\%$.
\end{enumerate}
Accordingly, our chatbot facilitators presented other collaborators' ideas with a similarity score below $2.0$ as dissimilar, above $2.0$ as similar, and above $3.0$ as very similar, which are filtered out as being considered repetitive ideas.

\section{Multi-armed bandit}
\label{appendix:mab}
We implemented a Multi-Armed Bandit (MAB) system for our adaptive facilitator. The MAB system selects an action (e.g., our chatbot's facilitator behaviors), receives a reward (e.g., users' rating on their own ideas), and selects the next action to try. 

\subsection{Upper confidence bound algorithm}
For the MAB, we used the Upper Confidence Bound (UCB) algorithm:
\begin{equation}
    A_t = \arg \max_{a}\left[Q_t(a) + c\sqrt\frac{\log t}{N_t(a)} \right]
\end{equation}
Where:
\begin{itemize}
    % \item $A$ = a set of actions (e.g., A = $\{$action $a$, $b$, $c$, ... , $n$$\}$)
    % \item $t$ = a time step of a trial that the MAB system performs.
    \item $A_t$ = action selected at trial $t$.
    \item $Q_t(a)$ = estimated reward of action $a$ at trial $t$.
    \item $N_t(a)$ = number of times that action $a$ has been selected up to trial $t$.
    \item $\log t$ = total number of trials that the MAB system has performed.
    \item $c$ = constant that controls the level of exploration.
\end{itemize}
\vspace{5mm}
The estimated reward ($Q_t(a)$) is computed with the following formula:
\begin{equation}
    Q_t(a) = m_t + \frac{r_t - m_t}{N_t(a)}
\end{equation}
Where:
\begin{itemize}
    % \item $r$ = reward
    \item $r_t$ = reward received at trial $t$.
    \item $m_t$ = the mean reward of action $a$ up to trial $t$.
\end{itemize}
\vspace{5mm}
Based on the expressions, the algorithm can be understood as; 
\begin{quote}
    `for action $a$, compute the estimated reward ($Q_t(a)$) and the uncertainty ($\sqrt\frac{\log{t}}{N_t(a)}$) for trial $t$. Compute them for all the actions and select the one ($A_t$) that has the highest upper confidence bound; the weighted sum of estimated reward and uncertainty ($\arg \max_{a}\left[Q_t(a) + c\sqrt\frac{\log t}{N_t(a)} \right]$)'.
\end{quote}

\section{Actions of the adaptive Facilitator}
\label{appendix:bandit_design}
For each idea generation and selection phase, we developed an MAB system according to the adaptive facilitator's behaviors.
During the idea generation phase, the adaptive facilitator has four actions as the combination of two inspirations and two ideation methods:
\begin{itemize}
    \item ``Here are other members' ideas \emph{similar} to yours... Can you propose \emph{any} idea?''
    \item ``Here are other members' ideas \emph{similar} to yours... Can you propose an \emph{improved} idea?''
    \item ``Here are other members' ideas \emph{dissimilar} to yours... Can you propose \emph{any} idea?''
    \item ``Here are other members' ideas \emph{dissimilar} to yours... Can you propose an \emph{improved} idea?''
\end{itemize}
% IG reward
Responding to an action, a user will either generate an idea or request a different inspiration if the user could not generate an idea based on the action.
When the user generates an idea, the chatbot asks the user to rate how helpful the idea is for achieving the design goal using 7-point Likert scale. Accordingly, the chatbot receives a reward between 1-7. When the user requests different inspirations, the chatbot automatically receives the lowest reward of 1.
% 
% Show me other answer
% 
This means that the more helpful ideas that the user generates, the higher reward the chatbot receives, which will make the chatbot try that action more than the other actions.

% IS action
During the idea selection phase, the notable ideas collected from the idea generation phase become the actions (Actions = [idea $1$, idea $2$, idea $3$,..., idea $n$]).
% IS reward
When the chatbot presents an idea, users rate the idea using the 7-point Likert scale. The chatbot will then compute how diverse users' opinions are and how many users have rated this idea, which is the standard error of the ratings (more details in Supplementary.C). For instance, if one idea has more diverse and a fewer number of ratings, its standard error will be higher. Accordingly, the chatbot will receive a higher reward and try to show that idea first to the other.

% We evaluated how our MAB systems would explore and exploit during each collaborative ideation phase in a simulated environment. We saw that the MAB chose the actions in a reasonable way, e.g. quickly identifying a suitable inspiration method (exploitation) but also changing the inspiration method if it became less beneficial (exploration). A discussion of the simulated evaluation is given in the Supplementary Material.

\section{Rewards for the adaptive facilitator's actions}
\label{appendix:bandit_reward}
During each collaborative ideation phase, our MAB system receives different rewards according to users' ratings:
\begin{itemize}
    \item \textbf{Idea generation}: $r_t$ = a user's rating of their own idea between 1 (very unhelpful) and 7 (very helpful). For instance, if a user;
    \begin{itemize}
        \item rates their idea `very helpful' at trial $t$, the MAB system receives a reward $r_t = 7$.
        \item rates their idea `very unhelpful' at trail $t$, the MAB system receives a reward $r_t = 1$.
    \end{itemize}
    \item \textbf{Idea selection}: $r_t$ = standard error ($SE$) of users' ratings between 1 (not interested at all) and 7 (very interested). For instance, idea A has received one user's rating $\{7 \}$. If the next user;
    \begin{itemize}
        \item rates idea A 'very interested' at trial $t$, the MAB system receives a reward $r_t =$ $SE$ of $\{7, 7\}$.
        \item rates idea A 'not interested at all' at trial $t$, the MAB system receives a reward $r_t =$ $SE$ of $\{7, 1\}$.
    \end{itemize}
\end{itemize}
The formula for the $SE$ is:
\begin{equation}
    r_t = \frac{\sigma(X_t)}{\sqrt{n(X_t)}}
\end{equation}
Where:
\begin{itemize}
    \item $X_t$ = set of user ratings for the proposed idea up to trial $t$.
    \item $\sigma(X_t)$ = standard deviation of $X_t$.
    \item $n(X_t)$ = number of samples in $X_t$.
\end{itemize}

\subsection{Simulated evaluation}
\begin{figure}[b!]
  \centering
  \includegraphics[width=\linewidth]{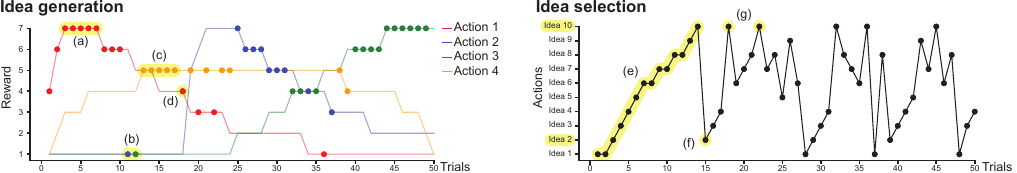}
  \caption{The evaluation of the MAB system in the simulated environment. We observed that the MAB system adaptively finds the most rewarding action during idea generation (left) and adaptively prioritizes the most uncertain ideas during idea selection (right).}
  \label{fig:mab_g_s}
\end{figure}

We evaluated how our MAB systems would explore and exploit during each collaborative ideation phase in a simulated environment. We created an environment that simulates a user's rating behavior during the idea generation phase (Figure~\ref{fig:mab_g_s}). We assume that a user's ideation performance (i.e., how well they could generate helpful ideas responding to each inspiration and ideation method) would change over time. For instance, users might generate more helpful ideas by building on similar ideas at the beginning. As they run out of ideas by thinking about a similar subject, they might generate ideas more easily from dissimilar ideas later on.
Accordingly, we predefined a set of rewards for four actions that would be most rewarding at the beginning, middle, end, or relatively lower than the other three actions throughout the trials.
Assuming that users would not generate ideas forever, we tested the MAB system for up to 50 trials.

The simulation showed that the MAB system mostly identified the best actions. For instance, in the first 10 trials (Figure~\ref{fig:mab_g_s}.a), the MAB system exploited Action 1 as it was the most rewarding. Then, responding to the decreased estimated reward of action 1 and increased uncertainty of the other actions, the MAB system explored Action 2 and 3 (Figure~\ref{fig:mab_g_s}.b), which gave the lowest rewards. In the following 5 trials (Figure~\ref{fig:mab_g_s}.c), the MAB system identified that Action 4 was most rewarding and exploited the action while exploring Action 1 (Figure~\ref{fig:mab_g_s}.d), which previously gave higher rewards than the other actions.

We created another environment that simulates a situation where users review and rate ideas during the idea selection phase (Figure~\ref{fig:mab_g_s}). We evaluated how the MAB system would adaptively re-order the ideas based on the ratings they receive. In the simulation, idea 2 and 10 had the highest standard error of the ratings. By selecting the idea 2 and 10, the chatbot would receive higher rewards.
The simulation showed that the MAB system was prioritizing the two ideas throughout the trials. In the first 14 trials, the MAB system was mostly exploring, collecting users' ratings on each idea.
Then, it identified that idea 2 was giving the highest reward, hence it prioritized idea 2 in the 15th trial. On the 18th and 22nd trials, the MAB system also prioritized idea 10 as it gave the higher reward. 

\end{document}